\documentclass{aa}

\usepackage{amsmath}

\usepackage{graphicx}
\usepackage{txfonts}

\begin{document}

\title{Laboratory-based grain-shape models for simulating dust infrared spectra}
\author{H. Mutschke\inst{1} \and  M. Min\inst{2} \and A. Tamanai\inst{1}}

\offprints{}

\institute{Astrophysikalisches Institut und Universit\"ats-Sternwarte (AIU), Schillerg\"a\ss chen 2-3, 07745 Jena, Germany
\and Astronomical Institute Anton Pannekoek, University of Amsterdam, Kruislaan 403, 1098 SJ Amsterdam, The Netherlands}

\date{Received <date> / Accepted <date>}

\abstract
{ Analysis of thermal dust emission spectra for dust mineralogy and physical grain properties depends on 
comparison spectra, which are either laboratory-measured infrared extinction spectra or calculated 
extinction cross sections based on certain grain models. Often, the agreement between these two kinds of spectra,  if available, is not yet satisfactory because of the strong influence of the grain morphology on the spectra. }
{ We investigate the ability of the statistical light-scattering model with a distribution of form factors (DFF) 
to reproduce measured infrared dust extinction spectra for particles that are small compared to the wavelength, i.e. in the size range of 1~$\mu$m and smaller. }
{ We take advantage of new experimental spectra measured for free particles dispersed in air with 
accompanying information on the grain morphology. For the calculations, we used DFFs that were derived 
for aggregates of spherical grains, as well as for compact grain shapes corresponding to Gaussian random 
spheres. In addition we used a fitting algorithm to obtain the best-fit DFFs for the various laboratory samples. 
In this way we can independently derive information on the shape of the grains from their infrared spectra. }
{ With the DFF model, we achieve an adequate fit of the experimental IR spectra. The differences in the 
IR band profiles between the spectra of particulates with different grain shapes are simply reflected by 
different DFFs. Irregular particle shapes require a DFF similar to that of a Gaussian Random Sphere with  $\sigma$=0.3, whereas roundish grain shapes are best fitted with that of a fractal aggregate of D$_f$=2.4-1.8. 
The fitted DFFs generally reproduce the measured spectral shapes quite well. For anisotropic materials, different 
DFFs are needed for the different crystallographic axes. The implications of this finding are discussed. }
{ The use of this model could be a step forward toward more realistic comparison data in infrared spectral 
analysis of thermal dust emission spectra, provided that these spectra are dominated by emission from submicron 
grains. }

\maketitle


\section{Introduction}
The analysis of mid-infrared dust emission spectra from stellar outflows, circumstellar disks, and 
other objects provides information about the dust mineralogy, grain sizes, and temperatures, hence, 
about physical and chemical conditions in the respective environments. The information about 
the dust grain properties is based on the lattice vibrational bands of the dust particles that mainly 
occur in the 10-50 $\mu$m wavelength range and that dominate the thermal emission of warm 
dust. Detailed analyses of, e.g., Spitzer IRS, and ground-based mid-infrared spectra including 
interferometric data (VLTI/MIDI) have been published in the last years for dust emission from 
accretion disks (e.g. van Boekel et al. \cite{vB04}), debris disks (e.g. Lisse et al. \cite{Lisse07}), 
and comets (e.g. Lisse et al. \cite{Lisse06}). 

The analyses are often performed by applying $\chi^2$ fits of a linear combination of either 
calculated or measured spectra of dust grains to the observed spectra. Both approaches have recently 
been improved by developing (1) new theoretical models (Min et al. \cite{Min05}, Min et al. \cite{Min06}) 
for calculating dust spectra and (2) a new experimental method for measuring spectra of 
particles dispersed in air (Tamanai et al. \cite{Tam06a}, Tamanai et al. \cite{Tam06b}), which 
avoids the influence of an embedding material on experimental infrared extinction spectra. 

These spectroscopic measurements of dust particles in aerosol also allow the investigation of the 
actual morphology of the aerosol particles by filtering and subsequent scanning electron microscope 
(SEM) imaging. Thus, the influence of morphological particle characteristics on the dust spectra can 
be studied in detail. Tamanai et al. (\cite{Tam06b}) and Tamanai et al. (\cite{Tam09b}) have reported 
a strong dependence of the measured profiles of the infrared bands on grain shape and agglomeration. 
Although the influence of the grain morphology on dust vibrational bands has been known for many 
years (Bohren \& Huffman \cite{BH83}), these effects have not been systematically studied in experiments 
before. 

\begin{figure*}
   \begin{minipage}[t]{9cm}
	\includegraphics[width=8.5cm]{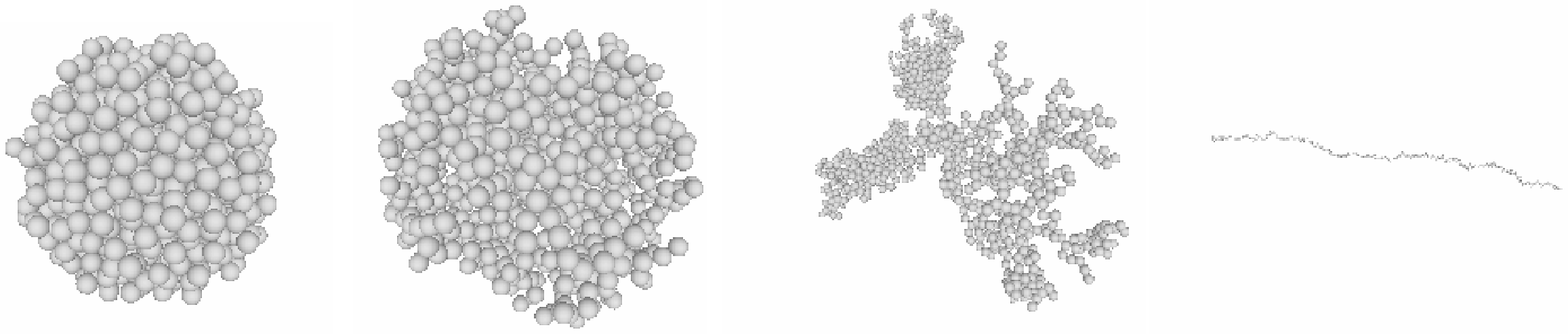}
	\hspace*{0.5cm}D$_f$=2.8\hspace{1.1cm}D$_f$=2.4\hspace{1.3cm}D$_f$=1.8\hspace{1.2cm}D$_f$=1.2
	\includegraphics[width=8.5cm]{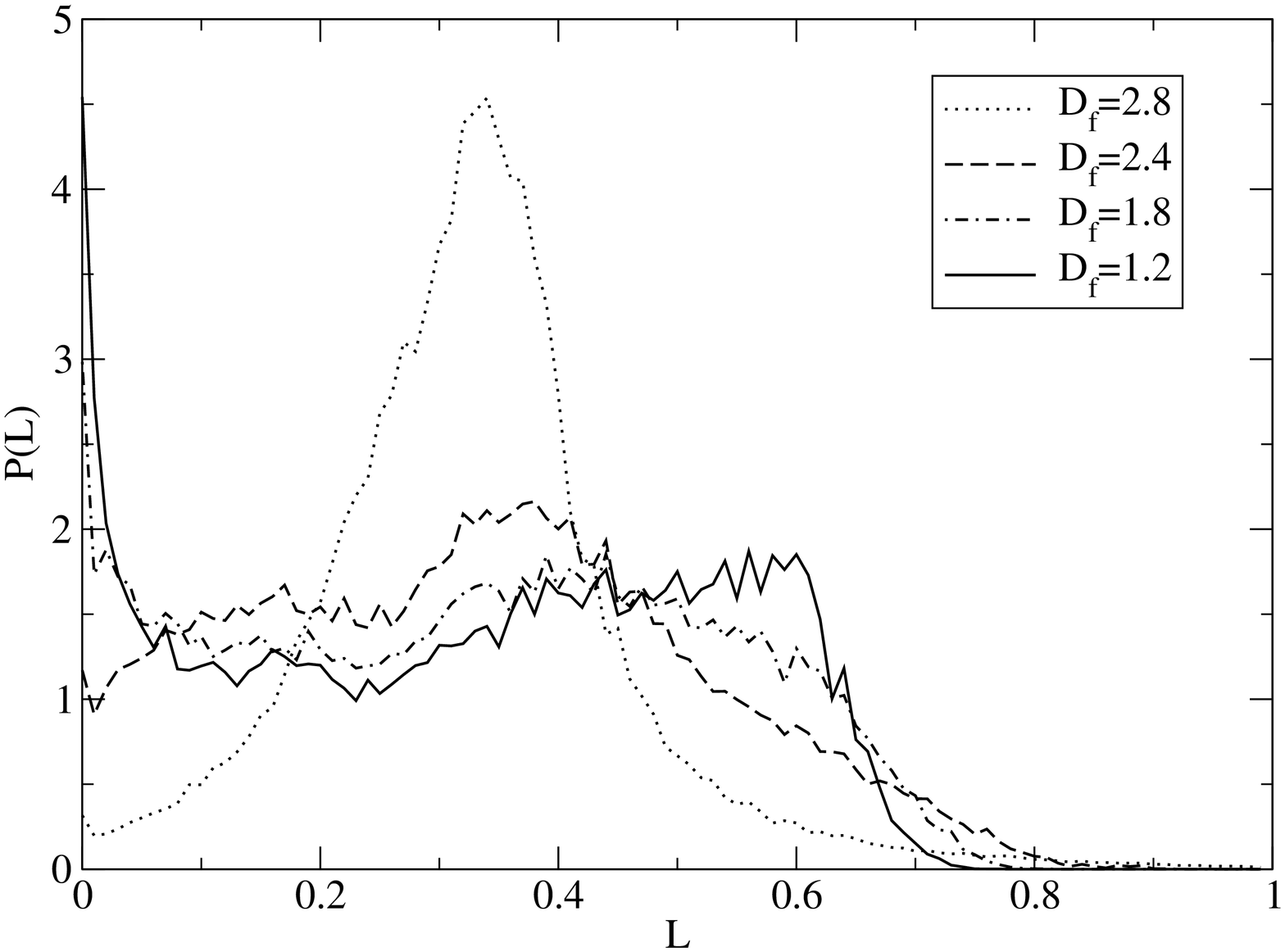}
   \end{minipage}
   \begin{minipage}[t]{9cm}
	\includegraphics[width=8.5cm]{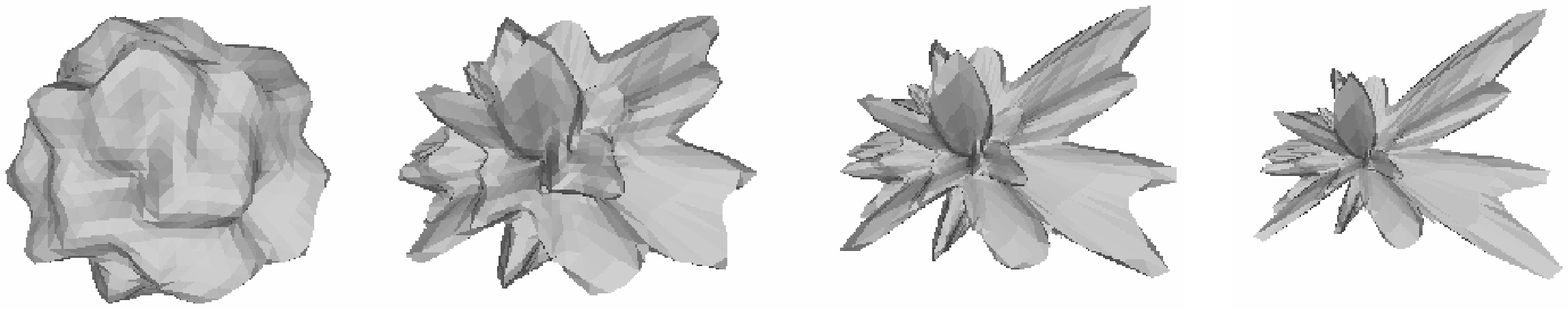}
	\hspace*{0.6cm}$\sigma$=0.1\hspace{1.3cm}$\sigma$=0.3\hspace{1.3cm}$\sigma$=0.5\hspace{1.3cm}$\sigma$=0.7
	\includegraphics[width=8.5cm]{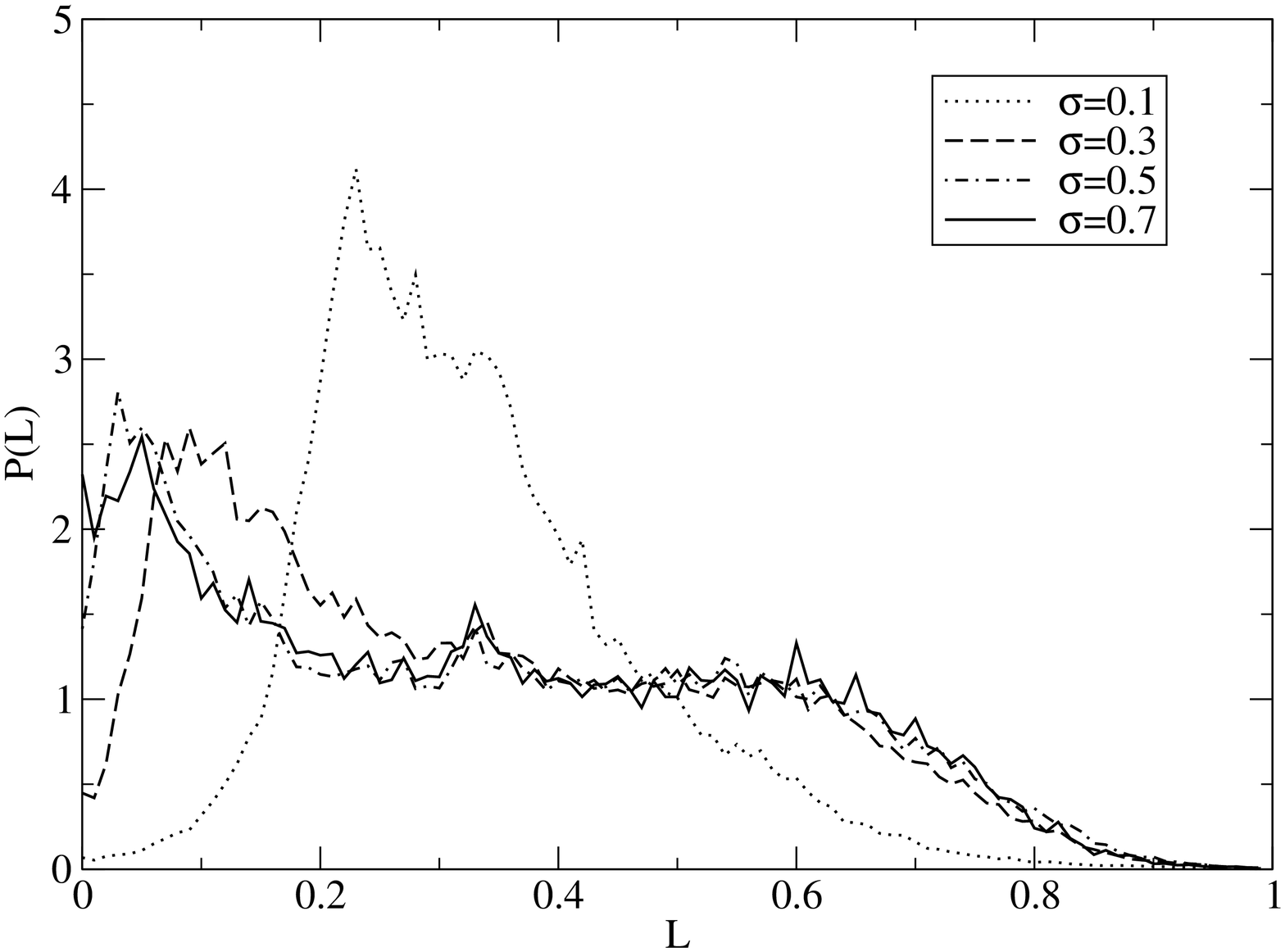}
   \end{minipage}
   \caption{Synthetic particle shapes and DFFs used in the modeling for a) aggregates of spherical 
   grains with different fractal dimensions D$_f$ and b) Gaussian Random Spheres with 
   different standard deviations $\sigma$.}
   \label{fig:dffs}
\end{figure*}

Furthermore, Tamanai et al. (\cite{Tam06b}) demonstrated that the band profiles predicted by the 
currently used models often differ considerably from the experimentally measured ones. 
The reason is that most of these models such as Mie theory, CDE (Bohren \& Huffman \cite{BH83}), 
and DHS (Min et al. \cite{Min05}) are not flexible enough to take morphological effects sufficiently 
into account. Since this is important for interpretating astronomical spectra, we aim at 
(a) a clarification of typical properties of measured infrared band profiles 
depending primarily on the grain shape, and (b) the investigation of the ability of an alternative 
theoretical model to take the shape effects into account.

For this purpose, we compare a selection of aerosol-measured spectra of oxides and silicates for two 
different classes of grain morphologies with calculated spectra using the theoretical approach of a 
form factor distribution (Min et al. \cite{Min06}). Both the experiments used here and the 
theory are restricted to particles small compared to the wavelength. For infrared wavelengths, this 
means a restriction to particles in the size range of approximately 1~$\mu$m and smaller. Consequently, 
the applicability of our results will be limited to interpreting emission (or absorption) 
spectra from cosmic environments where such grains are present. If larger grains provide a strong 
contribution to the emission cross section, the spectrum will be influenced by grain size effects that 
cannot be treated by the present approach. This is especially the case when growth or radiation effects 
have removed submicron grains, such as in evolved protoplanetary disks and debris disks around luminous 
stars. 

The theoretical model and the particulate materials are introduced in the next two sections, before 
we compare the experimental and the simulated spectra in Sect. 4. Section 5 summarizes the results.


\section{DFF model for spectra simulation}

The distribution of form factors model (DFF model, Min et al. \cite{Min06}) is a statistical 
approach valid for ensembles of particles that are small compared to the wavelength (Rayleigh limit). 
It describes the averaged extinction (absorption, emission) cross section $<$C$_{ext}$$>$ per unit 
volume of such a particulate by the integral over a continuous distribution of dipole polarizabilities 
depending on the dielectric function of the particulate material relative to its environment 
$\varepsilon=\varepsilon_p/\varepsilon_e$ and a depolarization factor or form factor L, which 
is defined in the range 0$\leq$L$\leq$1 (Eq.\,1). P(L) is the distribution of form factors over 
this range, which contains the information on the geometrical properties of the particles. 
\begin{equation}
\frac{\langle C_{ext}\rangle}{V}=\frac{2\pi}{\lambda}\int_0^1\frac{P(L)}{1/(\varepsilon-1)+L} dL
\end{equation}
Min et al. (\cite{Min06}) have demonstrated that a distribution P(L) (hereafter called 
the DFF) can be calculated for each particle shape, which gives a correct representation of 
the extinction cross section of that particle, provided that a spatial discretization with 
sufficient accuracy is possible. For a distribution of shapes, the individual DFFs have to be 
averaged. 

\begin{figure}
	\includegraphics[width=9cm]{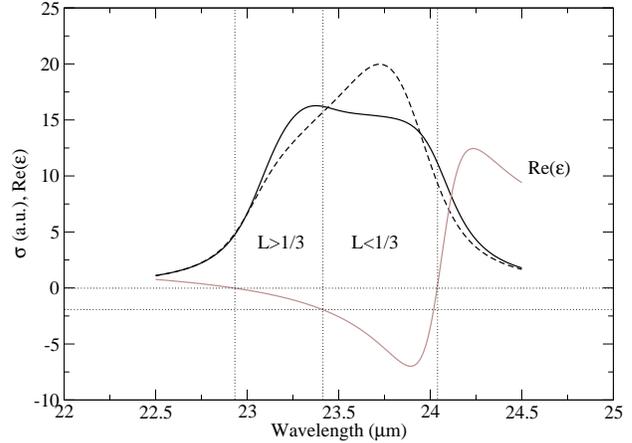}
	\caption{Extinction spectra of fractal aggregates with $D_f=2.4$ (solid black line) and Gaussian 
	random spheres with $\sigma=0.3$ (dashed line) calculated for a Lorentzian oscillator 
	type dielectric function. Re($\varepsilon$) is also shown (gray line). The vertical lines 
	indicate the wavelength range where Re($\varepsilon$)=1-1/L can be fulfilled and the position 
	of the sphere resonance (Re($\varepsilon$)=-2). }
   \label{fig:Lorentz}
\end{figure}

Another nice property of the DFF approach, which allows very intuitive understanding of the 
simulated spectra, is that major contributions to the integral occur when 
\begin{equation}
\varepsilon\sim1-1/L 
\end{equation}
(so-called surface modes, see Bohren \& Huffman \cite{BH83}). This is indeed the case in 
strong lattice vibration bands, where the real part of $\varepsilon$ takes a range of 
negative values (depending on wavelength), while the imaginary part is comparably small. 
Thus, relation (2) links the DFF directly to the extinction at corresponding spectral 
positions; e.g., for spherical grains (P(L)=$\delta$(L=1/3)), the extinction cross section 
spectrum will show a single resonance at the wavelength where $\varepsilon$=-2. 
Assuming a simple Lorentzian behavior of the dielectric function, the 
DFF at low L values is related to the strength of absorption at wavelengths longward 
of the sphere resonance (Re($\varepsilon$)$<$-2), whereas the DFF at larger L is related 
to absorption at shorter wavelengths (lower negative Re($\varepsilon$) values). We 
illustrate this in Fig.~\ref{fig:Lorentz}, and for a more precise treatment in the complex 
$\varepsilon$ plane see Min et al. (\cite{Min06}). 

The often used continuous distribution of ellipsoids model (CDE, Bohren \& Huffman \cite{BH83}) 
can be considered a special case of the DFF model with P(L)=2(1-L). Min et al. (\cite{Min06}) 
have calculated DFFs resulting from certain particle shapes for the cases of 
(a) aggregates of spherical grains with different fractal dimensions D$_f$ and of (b) particles 
in the form of Gaussian random spheres (GRS, Muinonen et al. \cite{Mui96}) with 
different standard deviations $\sigma$ of the surface modulation. We may note 
here that GRS grain shapes are used in scattering models for large grains as well 
(e.g. Volten et al. \cite{Volten01}). However, this is an entirely different 
(ray optics) approach, where the size distribution is also reflected by the 
(often extreme) GRS shape. In contrast, in our small-particle-limit approach 
the GRS model describes only the grain shape. 

The DFFs and illustrations of the particle morphologies are shown in Fig.\,\ref{fig:dffs}. 
In both cases, the DFFs broaden strongly with variation of the respective parameter, 
i.e. when going from compact to fluffy (low-dimension) aggregates and from sphere-like 
to extremely structured GRS. 

However, the details are different in the two series. While the ``main peak'' of the DFF 
shifts to higher L with decreasing D$_f$ in the case of the spherical grain aggregates, the opposite 
is the case for the GRS with increasing $\sigma$. For the extreme cases, namely the aggregate with 
D$_f$=1.2 and the GRS with $\sigma$=0.7, these main peaks are located at L=0.6 and L=0.05, respectively. 
Consequently, we can expect the two types of DFFs to produce band profiles peaking on 
opposite sides of the sphere resonance position (Re($\varepsilon$)$\sim$-2). Figure~\ref{fig:Lorentz} 
demonstrates this for less extreme geometries, i.e. an aggregate with D$_f$=2.4 and a GRS 
with $\sigma$=0.3. The short-wavelength peak of the aggregate particle spectrum 
(Re($\varepsilon$)$\sim$-1.5) corresponds to L$\sim$0.4, whereas the peak of the 
GRS spectrum (Re($\varepsilon$)$\sim$-6) corresponds to L$\sim$0.15. According to the DFF, 
it was expected to occur at L$\sim$0.1. The difference probably stems from the influence of 
Im($\varepsilon$), which increases towards longer wavelengths. 

For the aggregates of spherical grains, there is actually another peak in the DFF, 
which becomes stronger and stronger with lowering D$_f$, namely at L=0. This component of the 
DFF produces a long-wavelength shoulder in the band profile in addition to the 
short-wavelength peak (see Fig.~\ref{fig:Lorentz}). The L=0 component indicates two- or 
one-dimensional elongated, i.e. plate- or needle-like, structures. Agglomerates of spheres 
with point-like contacts are in principle not expected to produce such a component. Its presence 
is a consequence of the limited resolution of the discretization, which does not allow 
between point-like contacts and extended contact areas characteristic of e.g. coalesced particle 
aggregates to be distinguished (cf. Andersen et al. \cite{And06}). 
This is a direct effect of the method used to compute the DFFs (based on the discrete dipole 
approximation). However, coalescence of particles is often observed in real particulates, 
and its presence in the model distributions is not at all a shortcoming for their application, 
as we show. 

In this paper, we make use of these DFFs by comparing simulated infrared extinction spectra 
for such synthetic particle shapes with the measured spectra of real particulates with known 
(different) shapes. We consider particulates composed of spinel, corundum, and forsterite (see next 
section), for which the wavelength-dependent optical constants $\varepsilon_p$($\lambda$) are 
known from the literature. We use data by Fabian et al. (\cite{Fab01a}) for synthetic stoichiometric 
spinel, Querry et al. (\cite{Querry85}) for corundum, and Sogawa et al. (\cite{Sog06}) for forsterite. 
For the last two, the cross sections obtained for the individual crystal axes are averaged. 
For the environment, the dielectric constant $\varepsilon_e$ is unity in the simulation 
of aerosol spectra, 2.31 in cases when the particles are embedded into KBr matrices, 
and 3.03 for the case of embedding into a CsI matrix. 

\section{Experimental spectra}

\begin{figure*}[t]
   \begin{minipage}[t]{6cm}
	\includegraphics[width=5cm]{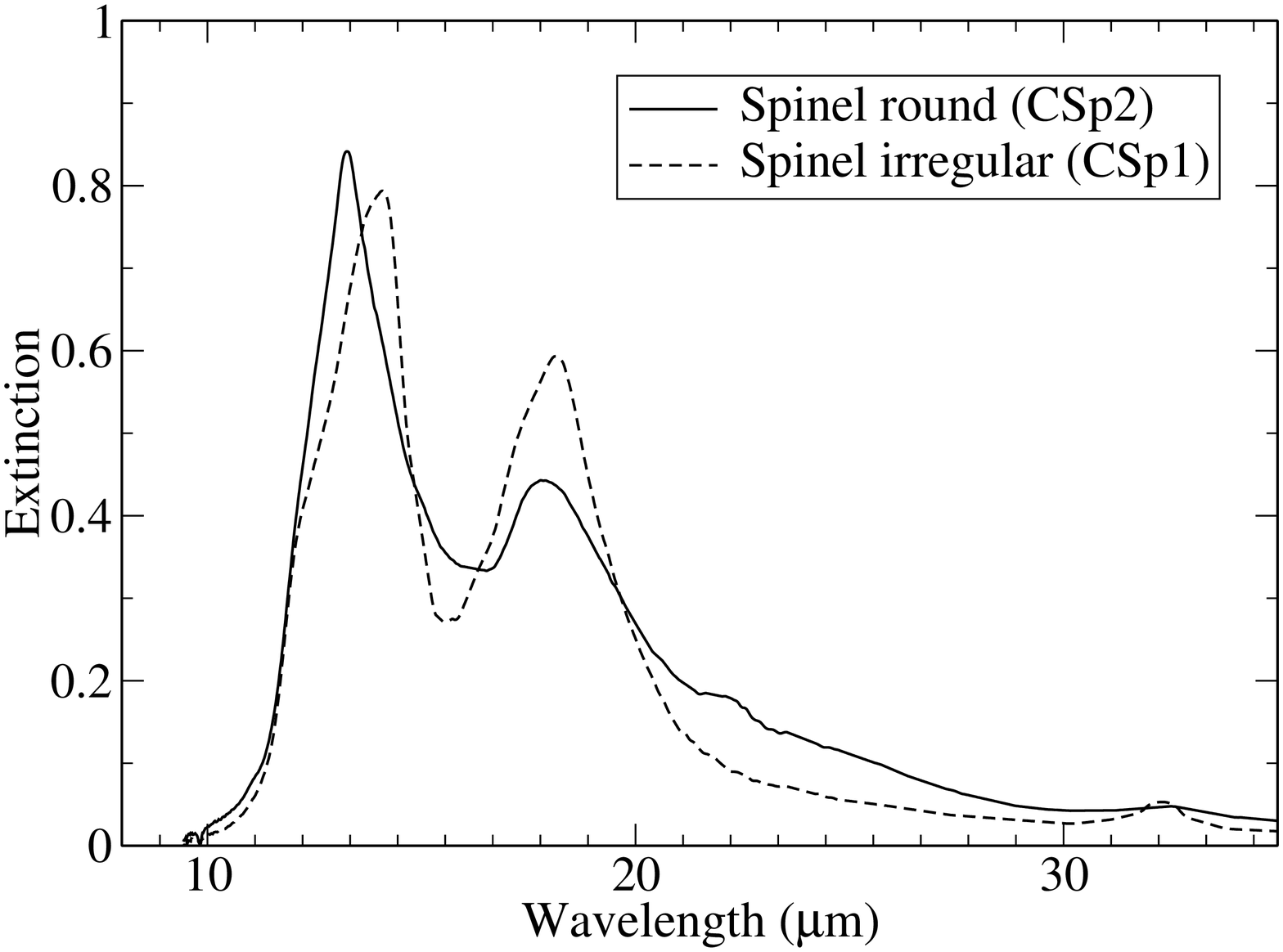}
	\includegraphics[width=5cm]{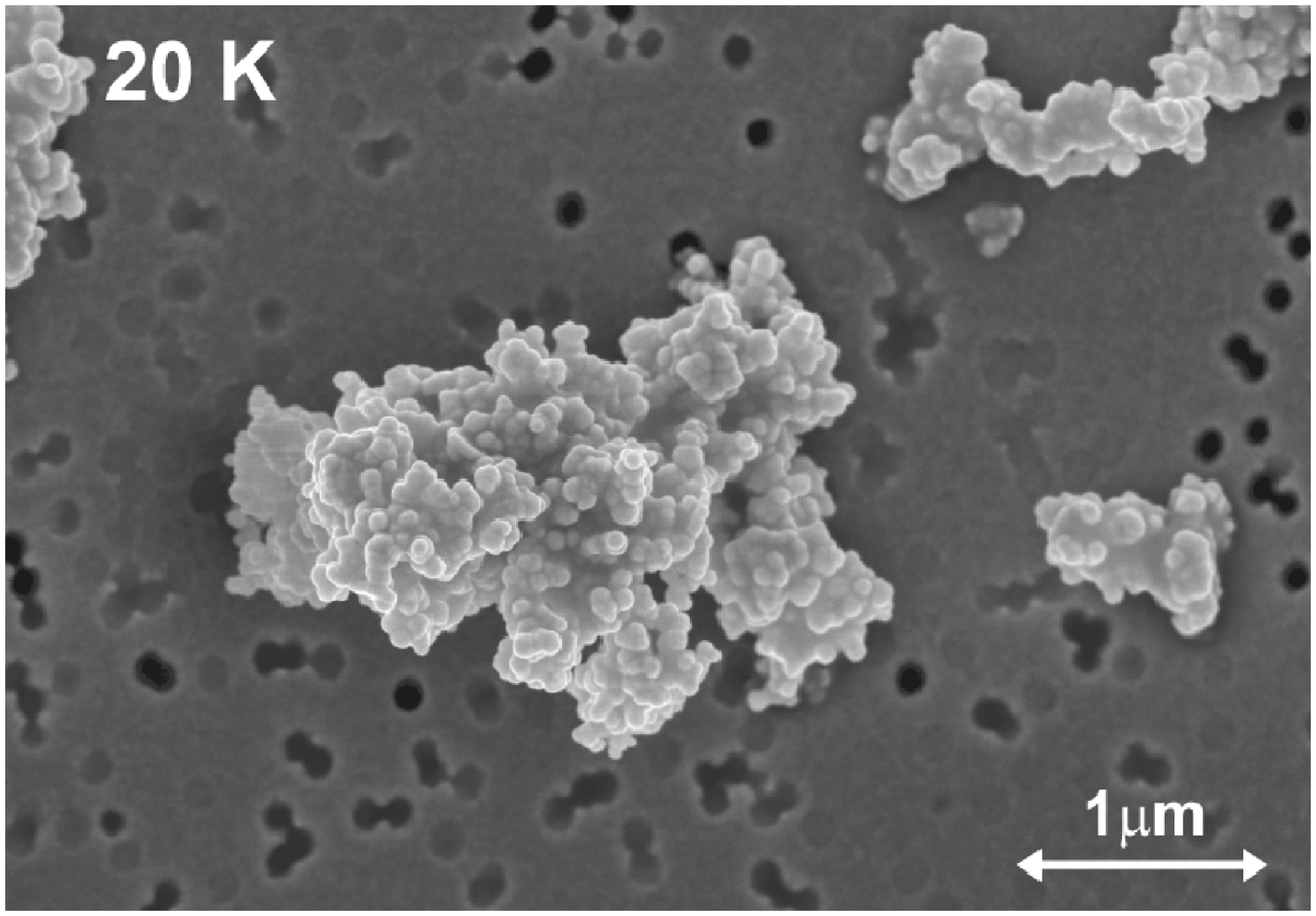}
	\includegraphics[width=5cm]{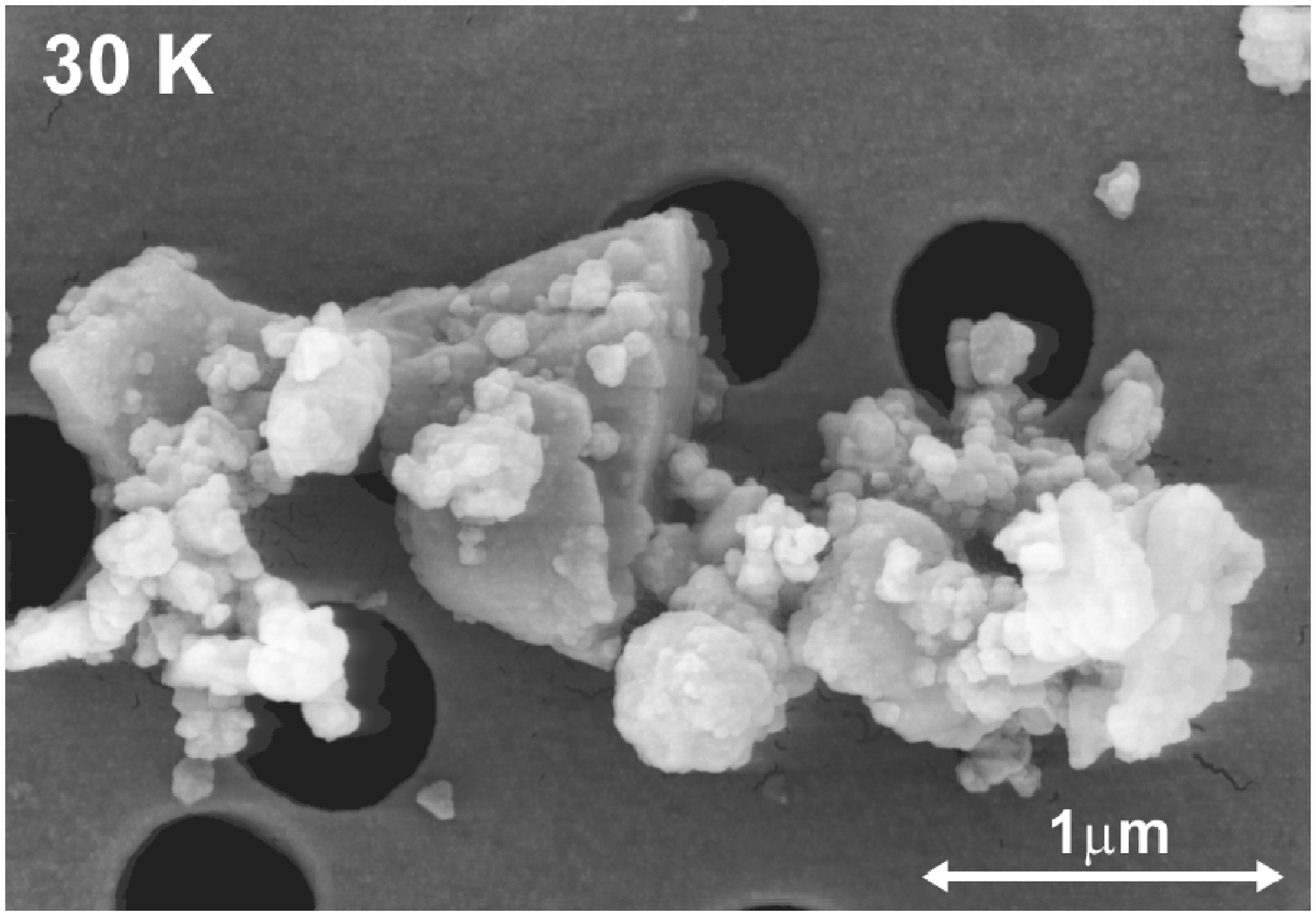}
   \end{minipage}
   \begin{minipage}[t]{6cm}
	\includegraphics[width=5cm]{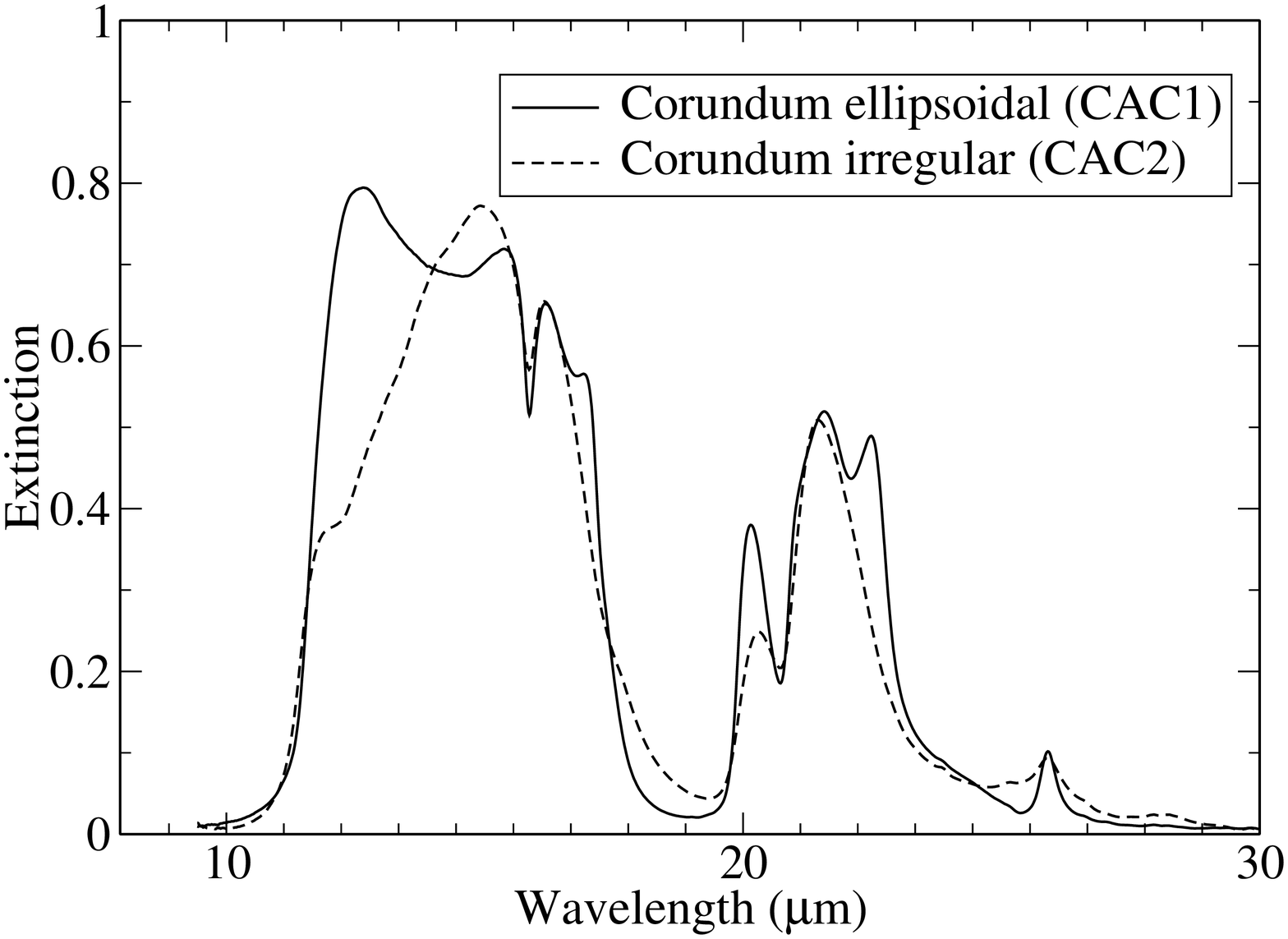}
	\includegraphics[width=5cm]{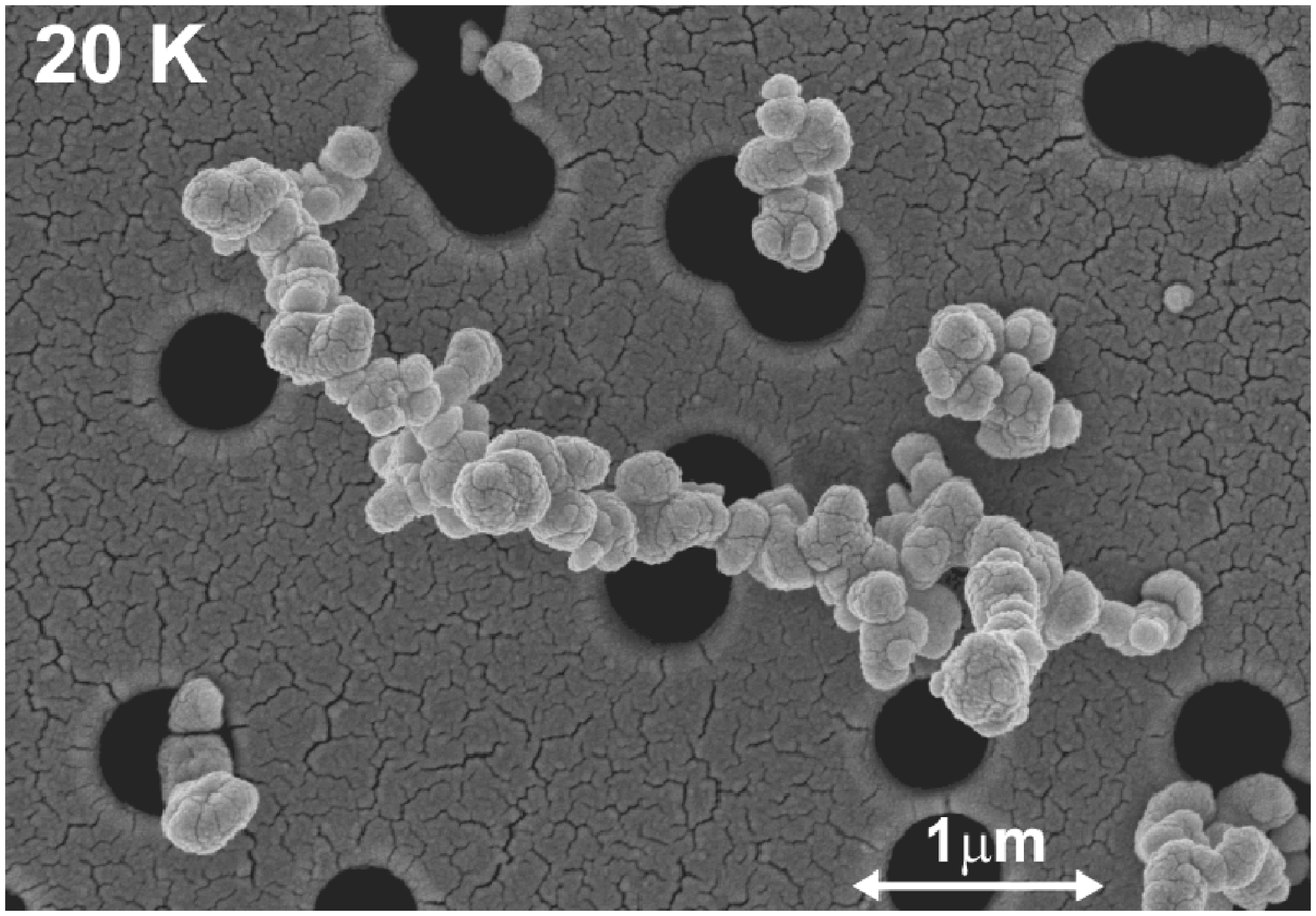}
 	\includegraphics[width=5cm]{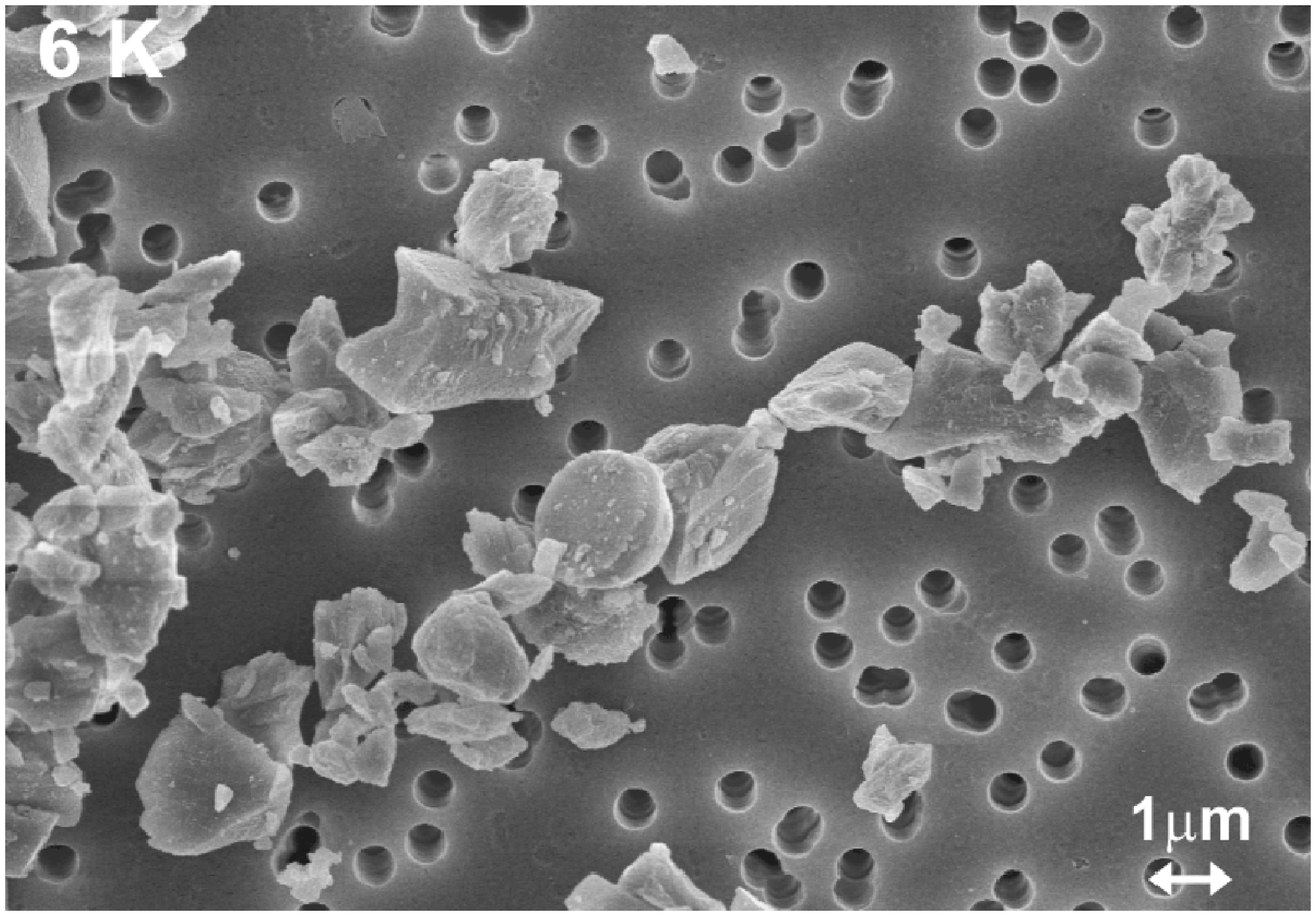}
   \end{minipage}
   \begin{minipage}[t]{6cm}
	\includegraphics[width=5cm]{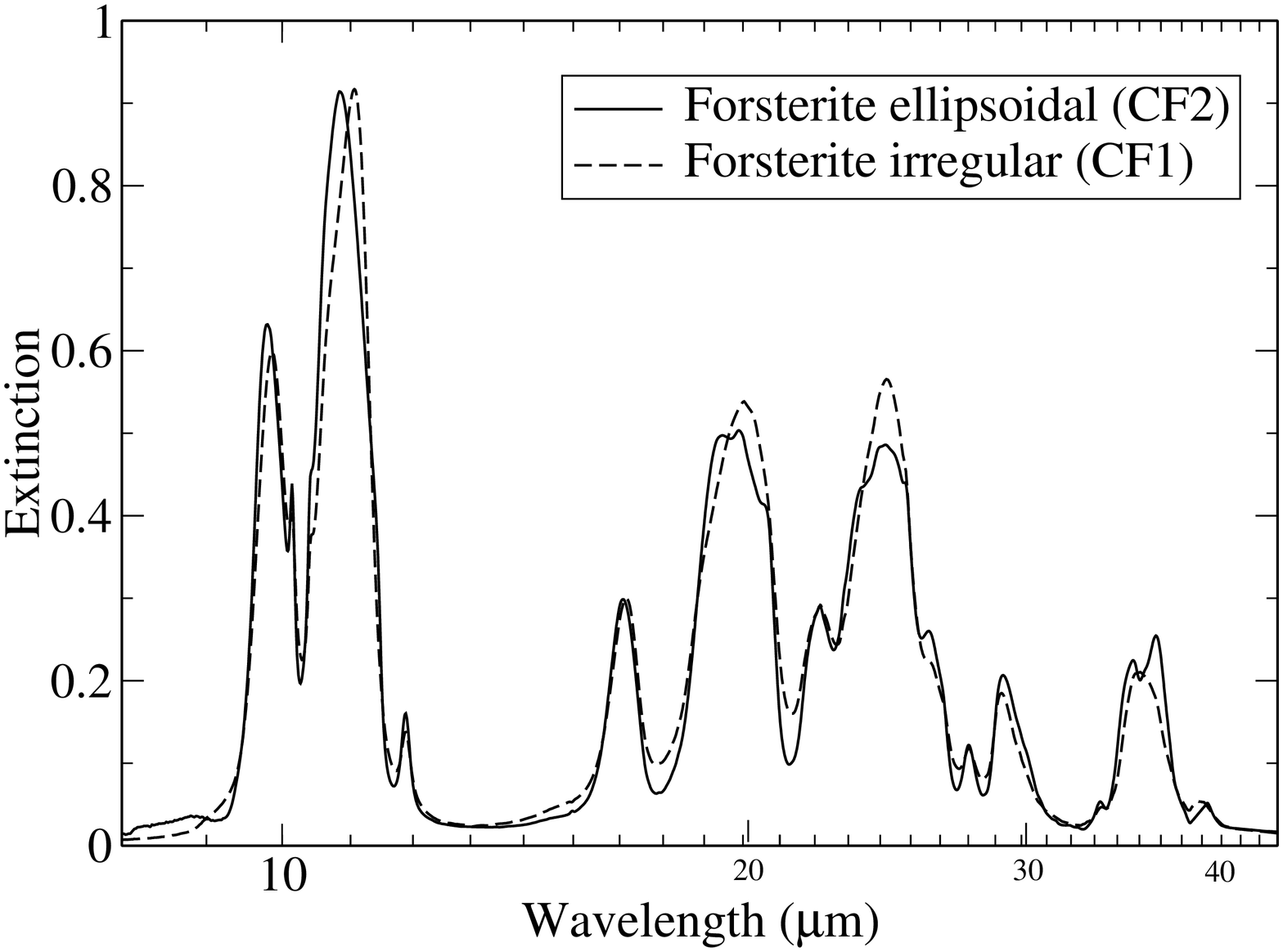}
	\includegraphics[width=5cm]{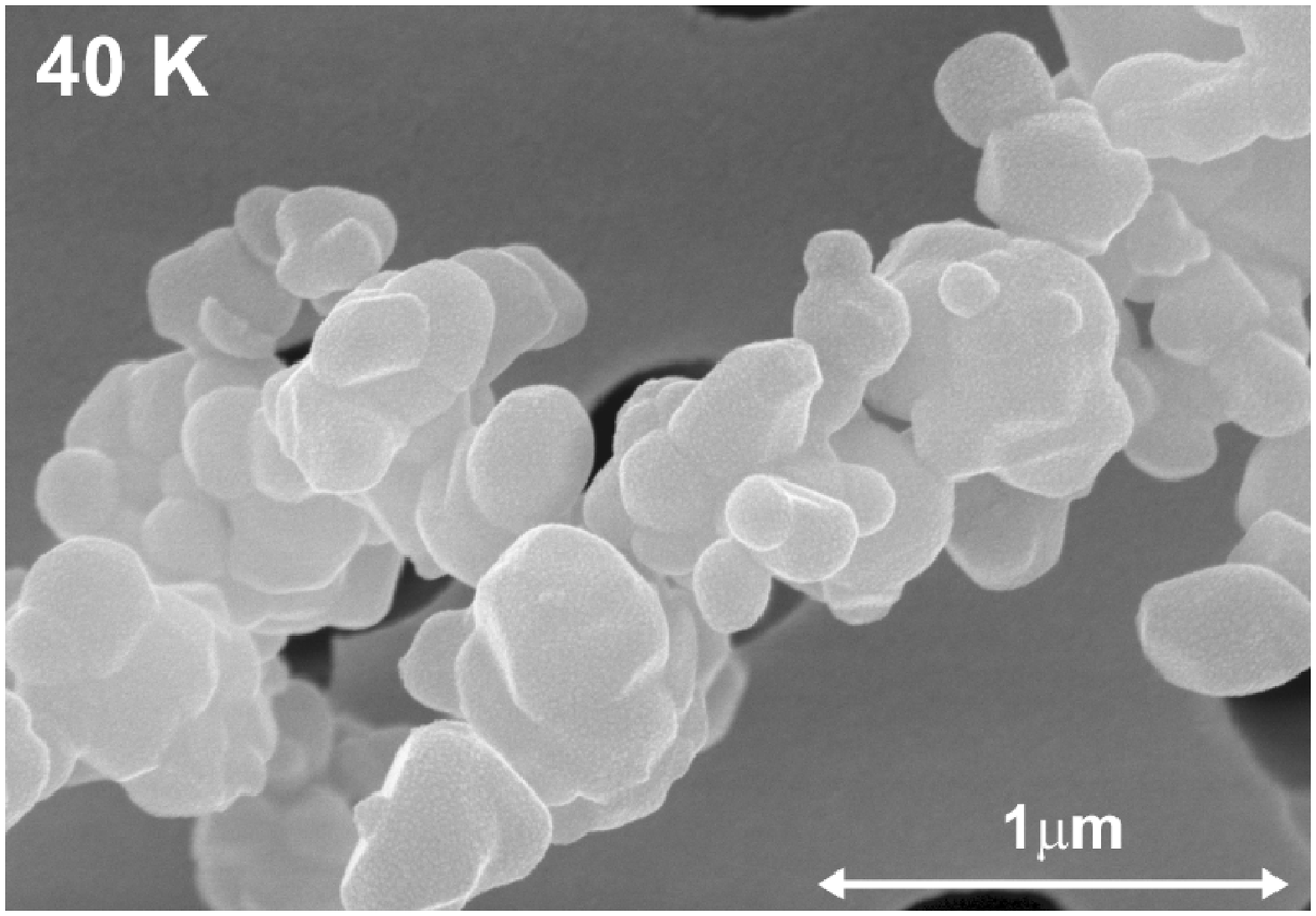}
 	\includegraphics[width=5cm]{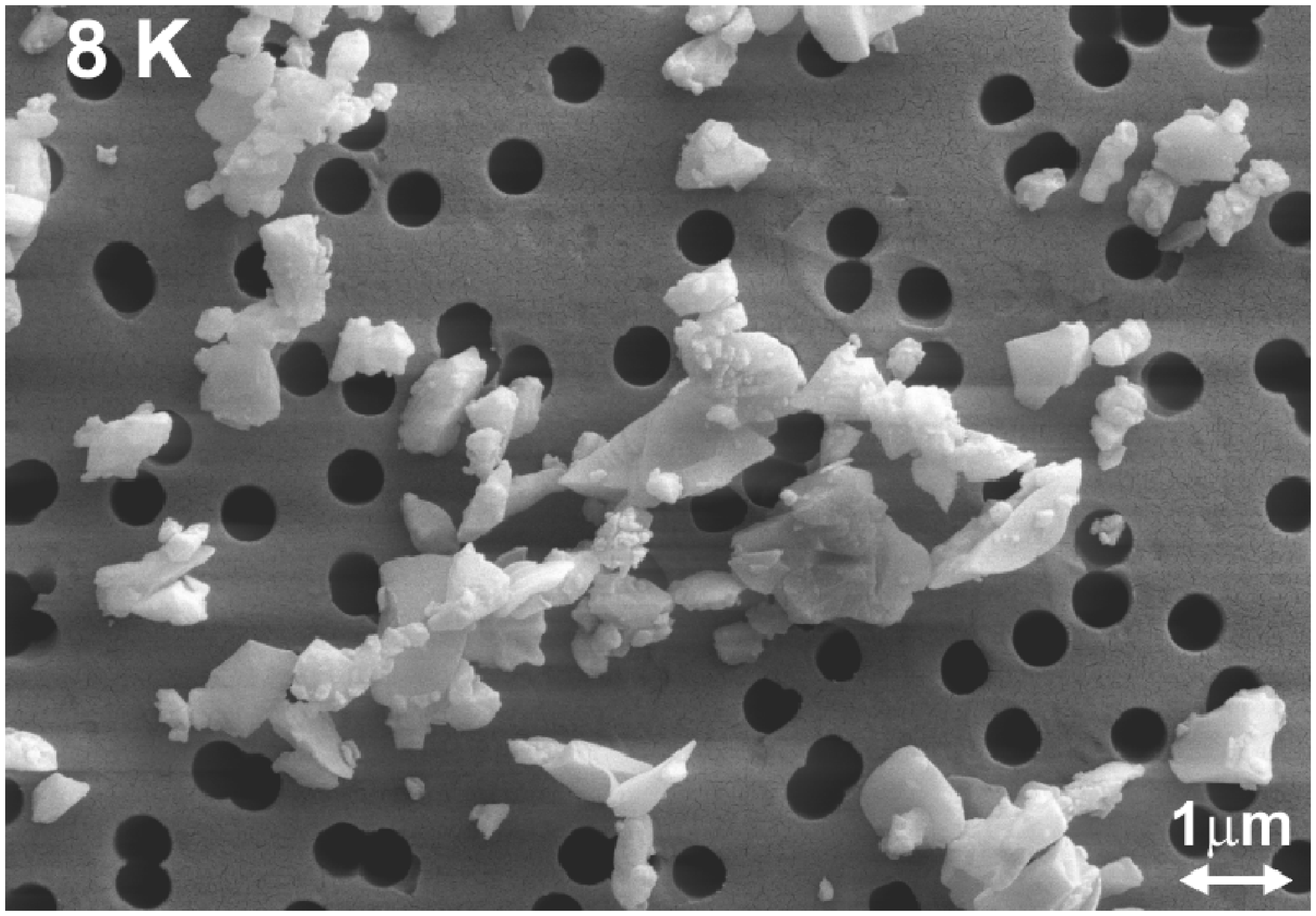}
   \end{minipage}
   \caption{Aerosol infrared extinction spectra and SEM micrographs of the 
   spinel (left), corundum (middle), and forsterite (right column) particulates. 
   The solid lines are the spectra of the particulates with roundish shapes, 
   which are shown in the images in the second row. The dashed lines correspond 
   to the spectra of the sharp-edged grains (images in bottom row). The dark 
   round filter holes in the image background are about 0.5\,$\mu$m in diameter 
   except for the image of the roundish spinel sample (upper left micrograph), 
   where they are 0.1\,$\mu$m in diameter. The numbers indicate SEM magnification. }
   \label{fig:shape}
\end{figure*}

The experimental setup for the measurement of infrared extinction spectra of dust particles dispersed in 
air (aerosol measurements) was described by Tamanai et al. (\cite{Tam06a,Tam06b,Tam09a,Tam09b}). 
For comparison with the simulations based on the DFF model, we have selected two samples of each 
of the three minerals spinel (MgAl$_2$O$_4$), corundum (Al$_2$O$_3$), and forsterite (Mg$_2$SiO$_4$), 
from these papers. The sample properties are listed in Table \ref{tab:Proben}. 

The two samples of each material have significantly different grain shapes. In general, one of the 
particulates can be characterized by a roundish shape or at least by having round edges, 
which is often the case for particles that have been condensed from the gas or liquid 
phase in a relatively fast process, whereas the other particulate is characterized by an irregular 
grain shape with sharp edges, which is typical of particulates produced by crushing larger pieces 
of material, but may also result from slow crystal growth. The spectra of the two classes of particulates 
differ significantly, as discussed by Tamanai et al. (\cite{Tam06b,Tam09b}). 
Correspondence with astronomically observed infrared dust emission bands has been found to be better 
in some cases for the roundish shapes, especially for the oxides in AGB star outflows (Tamanai et al. 
\cite{Tam09b}, see also Posch et al. \cite{Posch99} and Fabian et al. \cite{Fab01a}), but in other 
cases better for the irregular shapes, especially for forsterite (Tamanai et al. \cite{Tam06b}, 
see also Molster et al. \cite{Molster02}, Fabian et al. \cite{Fab01b}). Particle shapes and spectra 
are shown in Fig.\,\ref{fig:shape}. 

\begin{table*}
\centering
\caption{Properties of the samples.}
\label{tab:Proben}
\renewcommand{\footnoterule}{}
\centering
\begin{tabular}{|c|c|c|c|c|c|c|}
	\hline
Material & Chemical formula & Product info & Processing & Grain size & Grain shape & Reference\\
	\hline
Spinel	& MgAl$_2$O$_4$	& Aldrich & & $<$ 0.1 $\mu$m & round & T09b, CSp2\\
	&		& Alfa Aesar & Sedimentation & $<$ 1 $\mu$m & irregular & T09b, CSP1\\
Corundum & Al$_2$O$_3$ & Alfa Aesar & & $<$ 0.3 $\mu$m & ellipsoidal & T09b, CAC1\\
	&		& GC Jena & Sedimentation & $<$ 2 $\mu$m & irregular & T09b, CAC2\\
Forsterite& Mg$_2$SiO$_4$ & Marusu & & $<$ 0.3 $\mu$m & ellipsoidal & T06b, CF2\\
	&		& Alfa Aesar & Sedimentation & $<$ 1 $\mu$m & irregular & T06b, CF1\\
\hline
\end{tabular}
\end{table*}


\section{Results}

\subsection{Simulated spectra vs. measured aerosol spectra}

\begin{figure*}[t]
   \begin{minipage}[t]{8cm}
	\includegraphics[width=8cm]{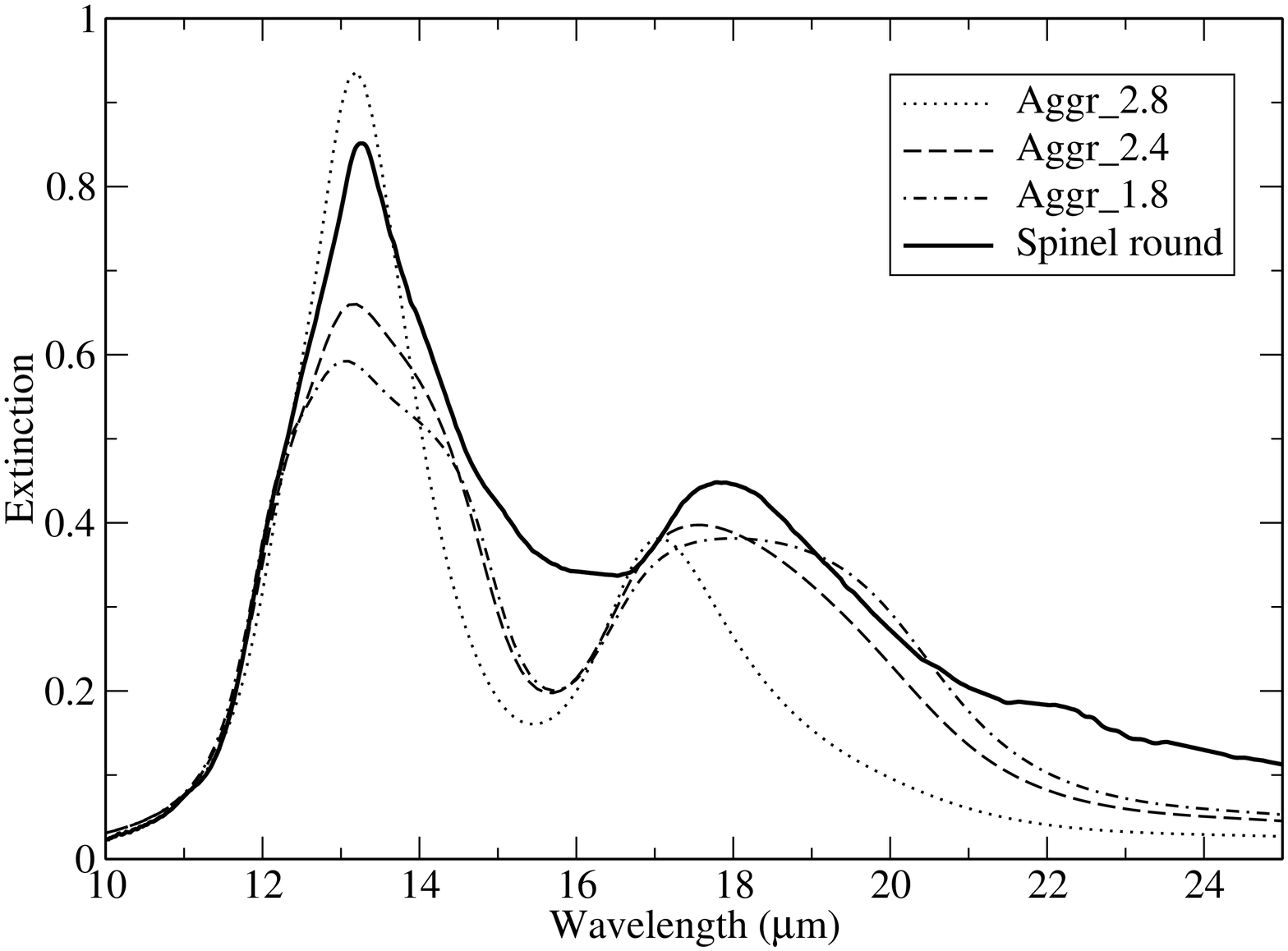}
	\includegraphics[width=8cm]{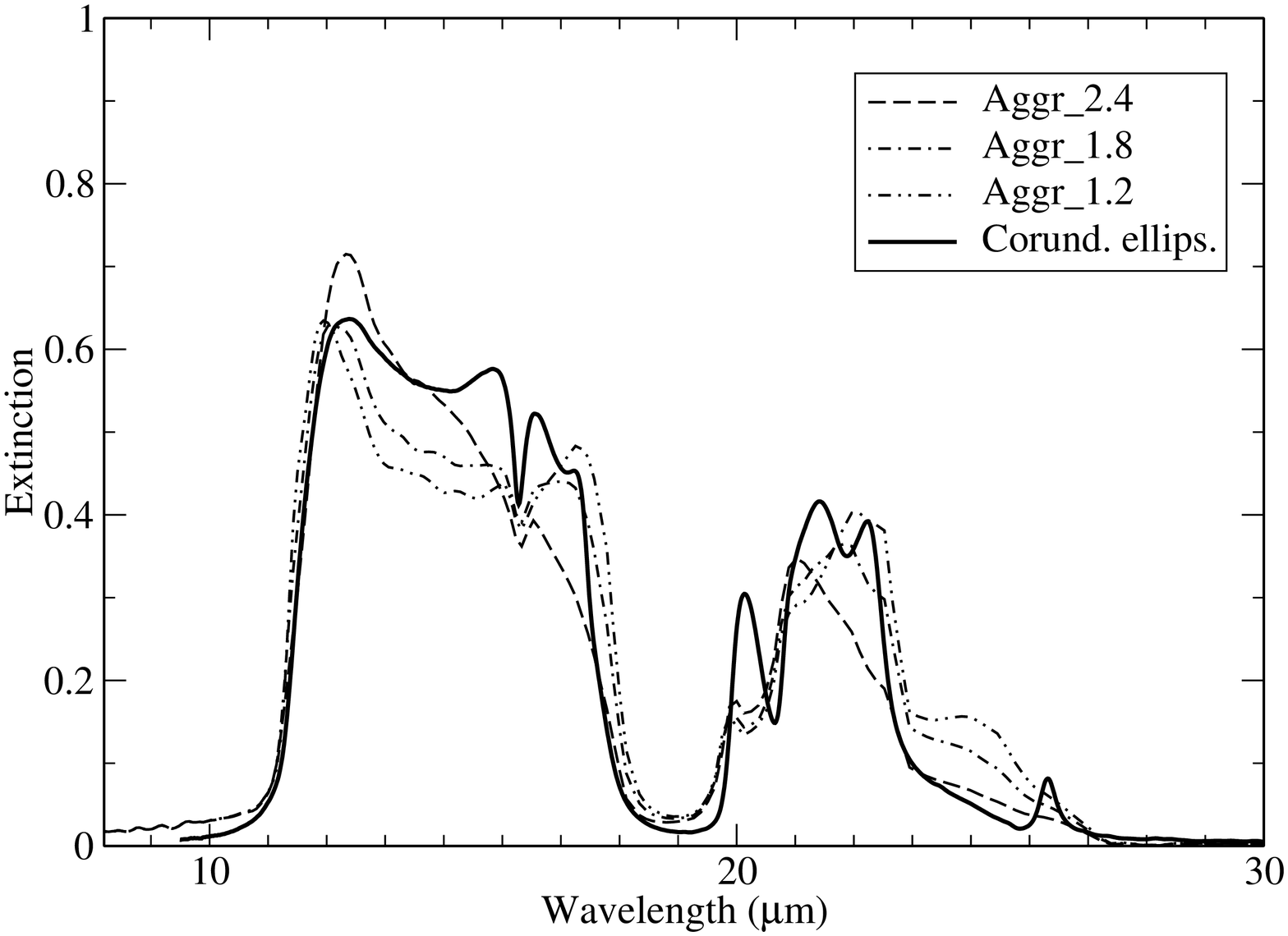}
	\includegraphics[width=8cm]{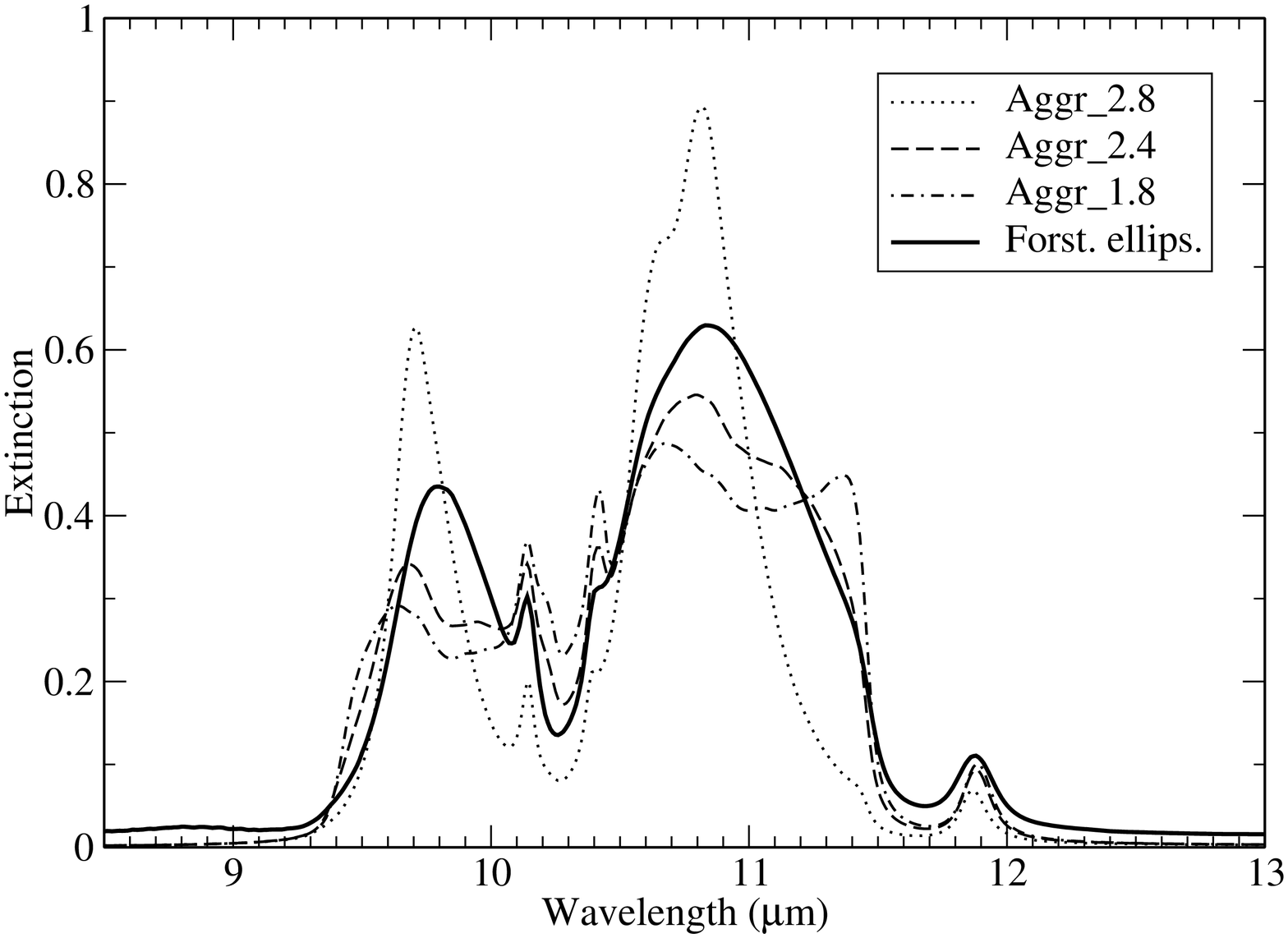}
	\includegraphics[width=8cm]{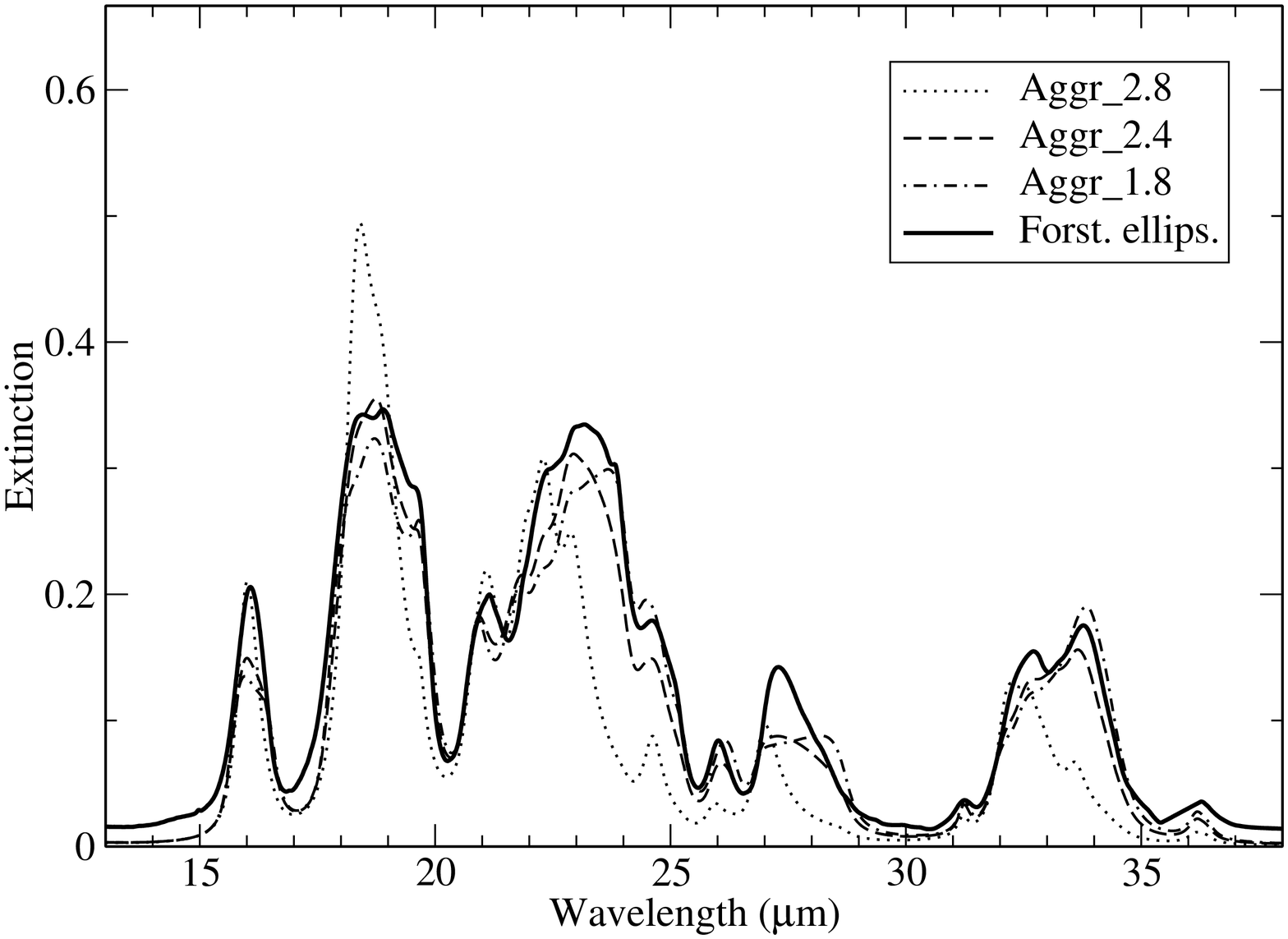}
   \end{minipage}
   \begin{minipage}[t]{8cm}
	\includegraphics[width=8cm]{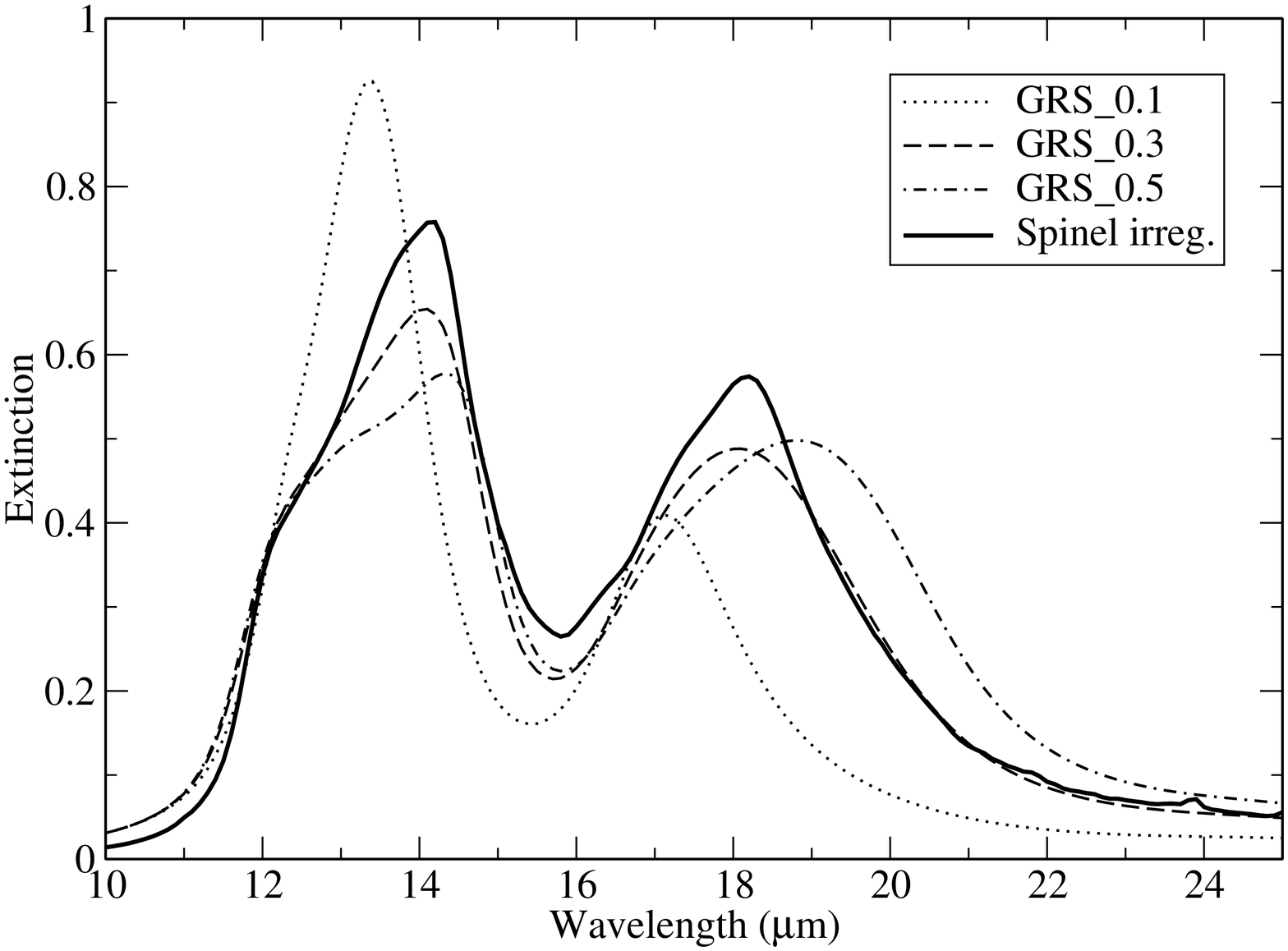}
	\includegraphics[width=8cm]{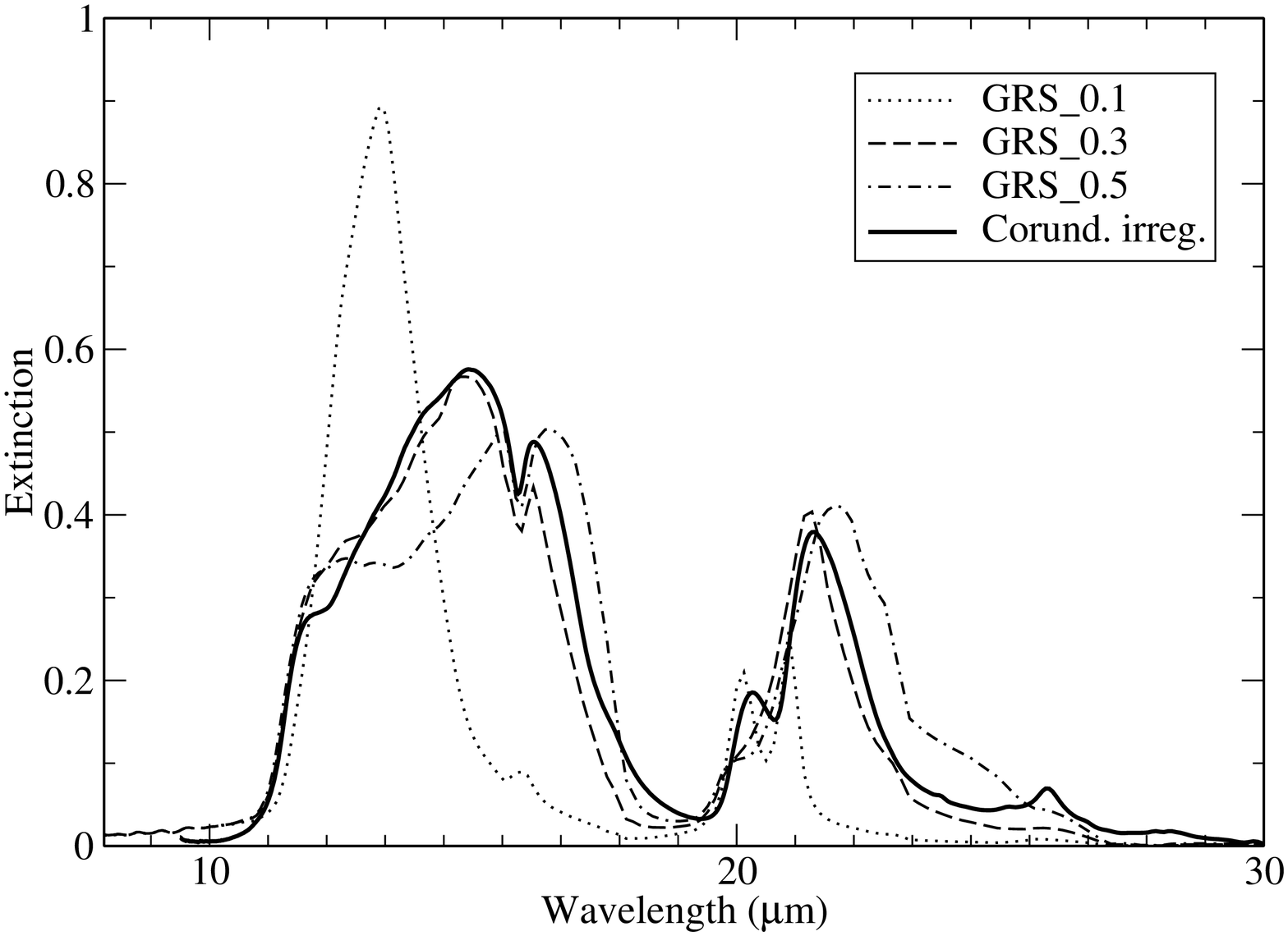}
	\includegraphics[width=8cm]{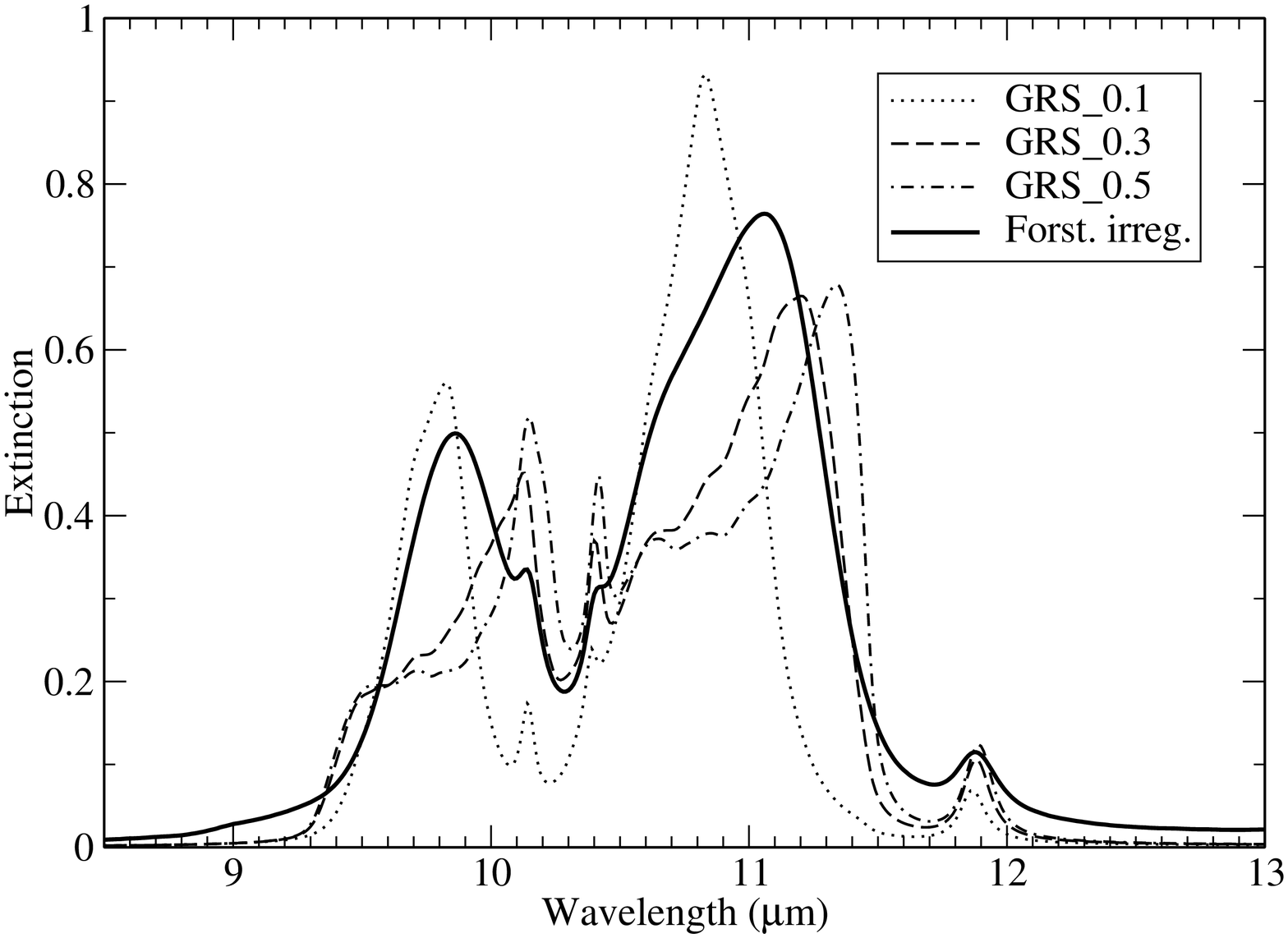}
	\includegraphics[width=8cm]{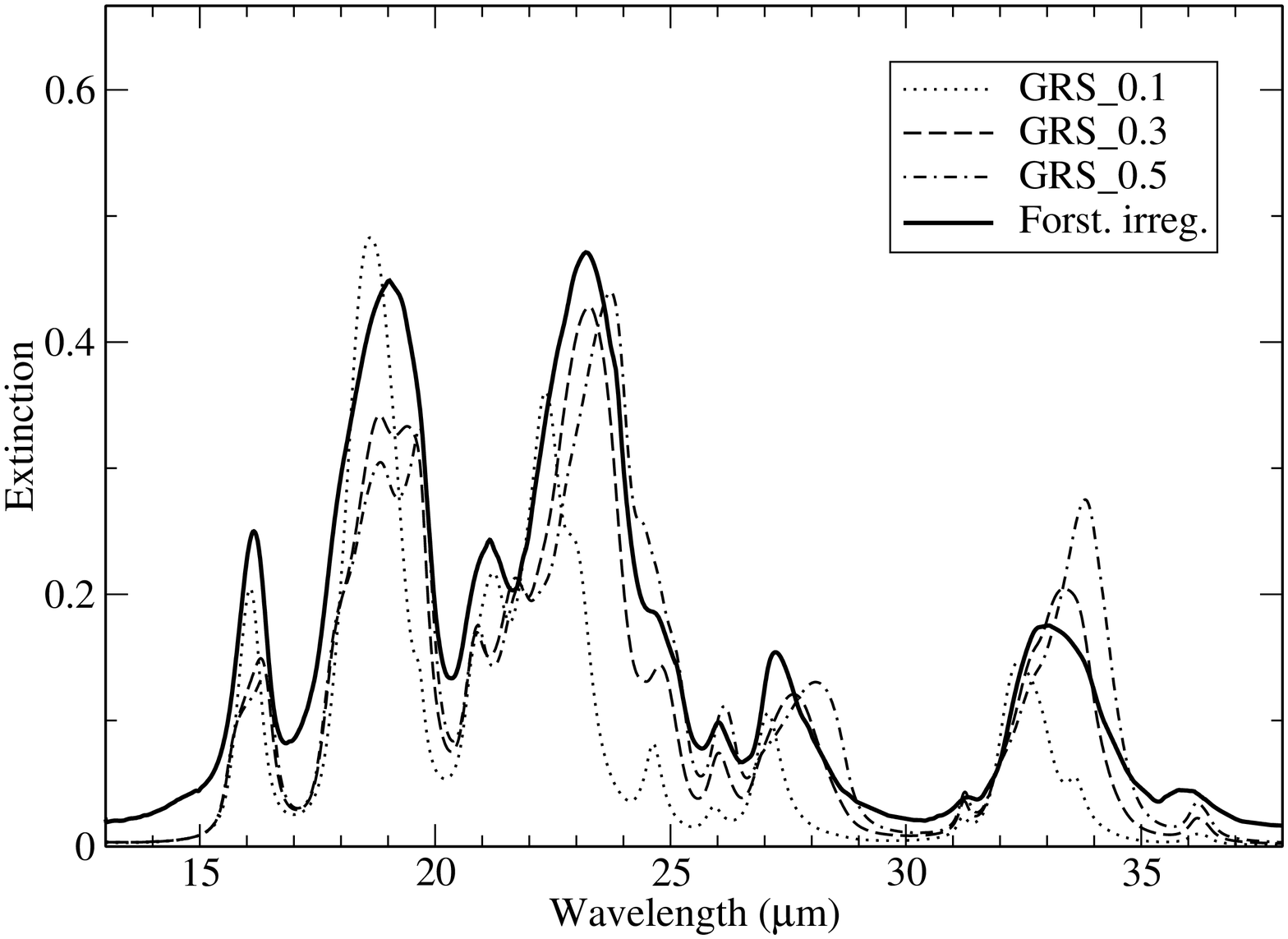}
   \end{minipage}
   \caption{Comparison of measured (solid lines) and simulated extinction spectra 
   for the roundish (left column) and irregular (right) particle shapes. DFF models 
   used in the simulations are given in the legends. The forsterite spectra have been 
   divided into two wavelength ranges.}
   \label{fig:comp}
\end{figure*}

Figure\,\ref{fig:comp} shows the comparison of the simulated spectra for spinel, corundum, 
and forsterite particles with the measured spectra. The left column compares the spectra 
calculated for the aggregates of spheres to the measured spectra of the roundish particulates. 
The simulated band profiles consist in most cases of peaks at short wavelengths and shoulders 
at longer wavelengths (compare Fig.~\ref{fig:Lorentz}). The latter strengthen with lowering 
fractal dimension of the aggregate, as do the L$\sim$0 components in the DFFs. The ``primary'' 
peak at shorter wavelengths, which weakens only between D$_f$=2.8 and 2.4, remains present 
while shifting to shorter wavelengths. This often leads to a ``rectangular'' band profile 
for low fractal dimensions, such as at D$_f$$\leq$1.8, which actually reproduces the 
measured spectrum of the corundum sample well. 
The comparison with the other simulated spectra demonstrates clearly that in the case of 
these particles a strong long-wavelength band component is definitely required to fit the 
measured profiles at about 17 $\mu$m and 22.5 $\mu$m. The finding that a fractal dimension 
of D$_f$=1.8 provides the best match corresponds nicely to the observed agglomeration 
state of the corundum powder, which is mainly chain-like (see Fig.\,\ref{fig:shape}). 
For the other samples of roundish grain shapes, the long-wavelength component is weaker but 
also present. For both the spinel and the forsterite samples, the D$_f$=2.4 spectrum fits 
quite well to the peaks in the measured spectra. 

In the case of spinel, the region in between the two bands and longward of 20\,$\mu$m shows 
an enhanced extinction that is not reproduced by any of the calculations. We interpret this 
feature as being caused by extinction from clumps of particles. Apparently, due to the small 
size of the individual grains, the clumps appear compact enough to act partially similar to 
large grains. According to Mie calculations, grains of about 5\,$\mu$m size already provide 
sufficient extinction (absorption {\it and} scattering) at these wavelengths. 

Moreover, we note that a few bands, such as the forsterite 10 $\mu$m and 27.5 $\mu$m and 
the corundum 20 $\mu$m bands, are not satisfactorily reproduced by the same models as the 
majority of the bands. This is discussed in Sect.\,\ref{sect:anis}. 

The right column compares the spectra of the irregular particulates with the simulations for 
the GRS of different surface modulation. As expected, the band profiles in these simulated 
spectra are dominated by peaks that shift continuously to larger wavelengths with an 
increasing standard deviation parameter. A shoulder remains at the short-wavelength edge 
of the profiles. The comparison with the measured spectra reveals excellent agreement 
for the $\sigma$=0.3 simulation in the cases of spinel and corundum (again apart from the 
$\lambda$=20\,$\mu$m band). For the forsterite, $\sigma$=0.3 is still the best-fitting calculated 
spectrum, even if it seems that a little lower $\sigma$ may have given a better match. 
This is especially true for the 10\,$\mu$m band, which we discuss separately 
in Sect.\,\ref{sect:anis}. 


\subsection{Fitting of DFFs and the problem of anisotropy}
\label{sect:anis}

An alternative way to compare theory and measurement is to fit DFFs to the measured spectra, 
hoping for a better reproduction of the measurements and aiming at the comparison of the 
obtained DFFs. We have developed an automatic fitting procedure that is able to handle 
anisotropic materials, i.e. contributions of up to three dielectric functions (in the case 
of forsterite, two for corundum) and even with the option of using different DFFs for each 
of these. In Sect.\,\ref{sect:media}, we show the result of this fitting for the irregular 
spinel. The fitting worked without problems for this isotropic material. 

In the previous section, we noted that a few bands in the spectra of the anisotropic 
materials show poor reproduction by the models, such as the 10 $\mu$m band of forsterite 
and the 20 $\mu$m band of corundum. For reproducing them, a DFF peaking strongly between 
L=0.2 and 0.25 (close to that of GRS 0.1) would work best, but it would not reproduce the 
other bands. These bands belong to vibrational modes along certain crystal axes, so 
introducing different DFFs for different crystal axes could solve these problems. 

\begin{figure}
	\includegraphics[width=9cm]{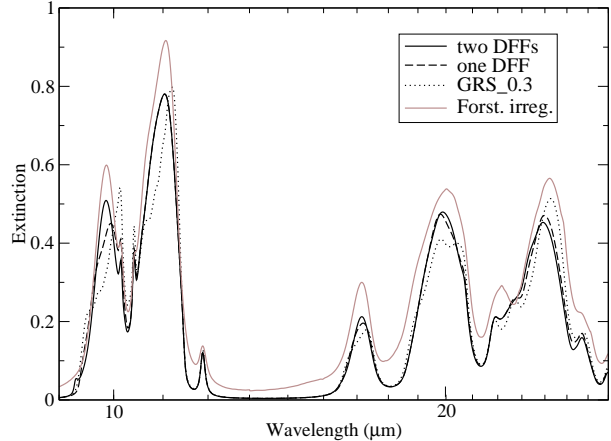}
\caption{Comparison of fits to the measured spectrum (gray line) of irregular forsterite 
particles using a single DFF for the contributions of all crystal axes (dashed line) and 
allowing two DFFs, i.e. a separate DFF for the contribution of the vibrational modes 
along the crystallographic x-axis (solid line). The DFFs obtained by the two-DFF fitting 
are plotted in Fig.\,\ref{fig:comp_dffs_irr}. For comparison, the simulation using the 
GRS model with $\sigma$=0.3 is also shown (dotted line).}
\label{fig:aj_fit}
\end{figure}

In general, deviations in the DFFs along different crystal axes are not an unlikely option, 
because the grains could be primarily elongated along certain crystal axes. 
Such grain shapes can be produced both by growth and shattering processes. 
Allowing for different DFFs for the different crystal axes lead to a satisfactory 
reproduction of the problematic bands both for forsterite and for corundum. 
Figure\,\ref{fig:aj_fit} shows as an example the fits for the irregularly shaped 
forsterite with (a) only one DFF for all axes and (b) a separate DFF for the 
contributions of the vibrations along the crystallographic x-axis. The 
better match in the 10\,$\mu$m band is clearly visible. 

\begin{figure}
	\includegraphics[width=9cm]{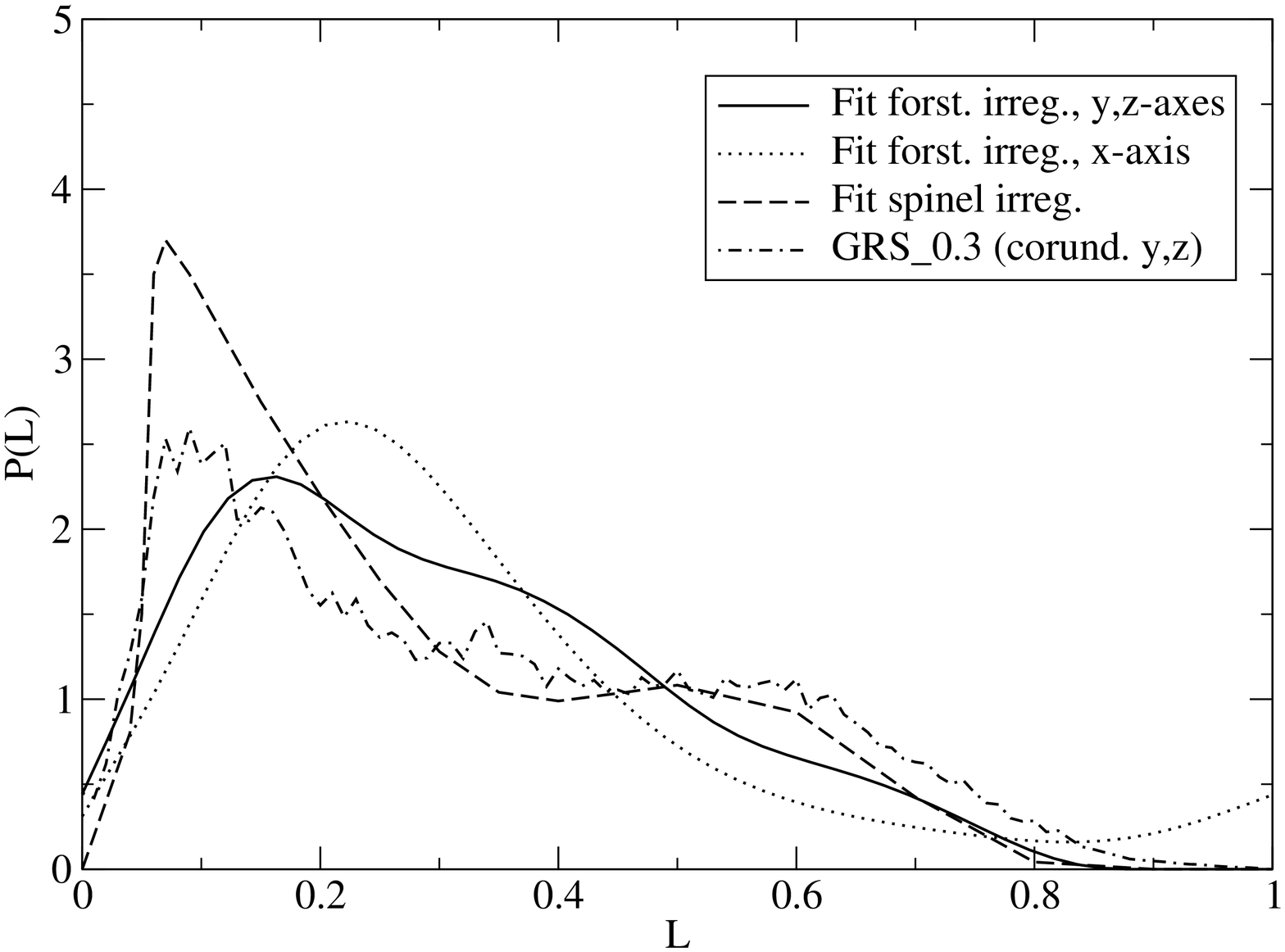}
\caption{Comparison of the fitted DFFs for irregular forsterite particles and irregular 
spinel particles (dashed line) with the GRS model with $\sigma$=0.3 (dash-dotted), which 
provides an excellent fit for the spectrum of the irregular corundum particles. }
\label{fig:comp_dffs_irr}
\end{figure}

However, the DFFs for the different axes needed to be very different. The DFF required 
by the problematic bands were always similar, with a narrow peak in the L=0.2-0.3 range. 
This changed only slightly for roundish grains, even if the main DFF peaked then on much 
higher L values. In Fig.\,\ref{fig:comp_dffs_irr}, the two fitted DFFs are shown for the 
irregular forsterite particles (see Fig.\,\ref{fig:aj_fit}), where the difference is still 
quite moderate. The GRS model with $\sigma$=0.3, however, represents the case well for 
the y and z-axis DFF for the irregular corundum particles. Here, the difference from the 
x-axis DFF is very strong. Especially for the roundish grains, such extreme differences 
between the crystal axes and the nondependence on the grain-shape seemed unlikely. In some 
bands, we noted that the poor fit introduced by the additional DFF had to be compensated 
for by the other components. 

This pointed to a different reason for the mismatch in some bands, which turned out 
to be an incorrectness in the model, namely in the treatment of the anisotropy of 
the materials. The simple averaging of the absorption cross sections calculated for 
the crystallographic axes is correct for spheres that are small compared to the wavelengths 
and for ellipsoids, if the ellipsoid principal axes are associated with the crystal 
axes. When assuming complicated shapes, such as the GRS or aggregates, however, 
it is clear that this condition is not maintained. Here, the influence of the 
polarization along one axis on the excitation along other axes has to be considered. 

This can be demonstrated by means of the discrete dipole approximation (DDA) model. 
The DDA allows one to take the full tensor nature of the refractive 
index into account. We will demonstrate the effects of this in a separate paper 
(Min et al. in prep). 


\subsection{Reproduction of KBr and CsI spectra}
\label{sect:media}
An interesting question is whether the DFF models can predict the influence of embedding 
media on the spectra of particulates. Tamanai et al. (\cite{Tam06b,Tam09b}) demonstrated 
the strong influence of potassium bromide and cesium iodide matrices used in the conventional 
pellet technique onto IR band profiles of embedded particulates. Within the DFF model, the 
embedding medium is taken into account simply via the relative dielectric constant $\varepsilon=\varepsilon_p/\varepsilon_e$.

\begin{figure}
	\includegraphics[width=9cm]{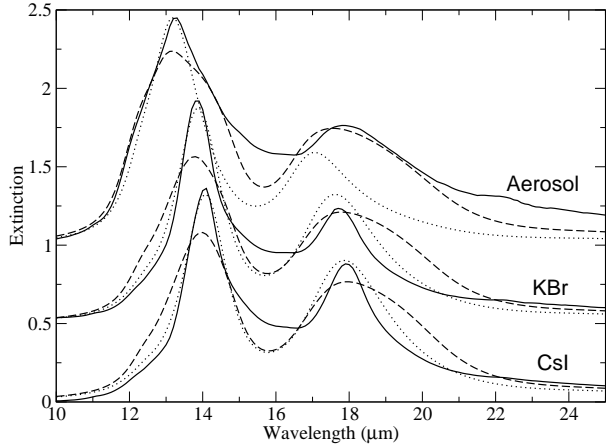}
\caption{Comparison of measured (solid lines) compared to simulated extinction spectra
for roundish spinel particles embedded in different media. The DFF models used are the
aggregates of spherical grains with D$_f$=2.8 (dotted lines) and D$_f$=2.4 (dashed lines).}
\label{fig:media_round}
\end{figure}

\begin{figure}
	\includegraphics[width=9cm]{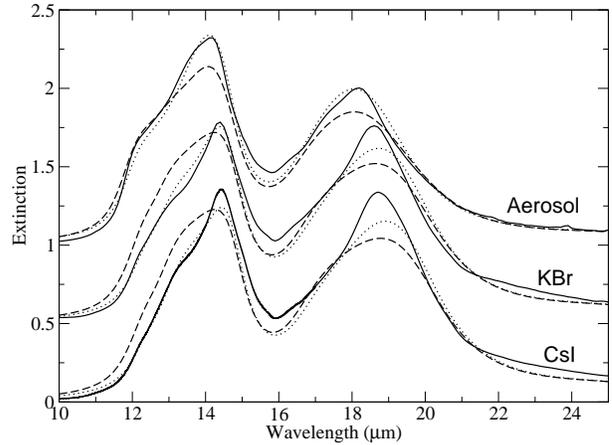}
\caption{Same as Fig.\,\ref{fig:media_round} for irregular spinel particles. The DFF models 
used here are the Gaussian random spheres with $\sigma$=0.3 (dashed lines) and a fitted DFF 
(see Fig.\,\ref{fig:comp_dffs_irr}).}
\label{fig:media_irr}
\end{figure}

Figure\,\ref{fig:media_round} and \ref{fig:media_irr} show simulated spectra for spinel 
using identical DFFs and different dielectric constants of the embedding medium 
($\varepsilon_e$=1 for aerosol, 2.31 for KBr, and 3.03 for CsI) compared to corresponding 
measured spectra. For the irregularly shaped particles (Fig.\,\ref{fig:media_round}), the 
DFF for GRS with $\sigma$=0.3 and a fitted DFF were used. The comparison shows that 
the former, which reproduces the aerosol measurements satisfactorily, gives a still quite 
imprecise representation of the measured spectra for embedded particles. However, this 
does not mean that the model fails to treat the influence of the embedding, because some 
adjustment of the DFF (see Fig.\,\ref{fig:comp_dffs_irr}) leads to a nearly perfect fit 
of all the spectra measured in different matrices.

Unfortunately, this is not the case for the roundish particles (Fig.\,\ref{fig:media_round}, 
see especially the 18\,$\mu$m band). Here, the embedded particles are best matched by a 
DFF strongly dominated by the spherical character of the single grains or characterized 
by very compact aggregation, whereas for the aerosol spectra we had to use a DFF calculated 
for a lower fractal dimension of the aggregates. If this indeed reflects different morphologies 
of the spinel particles in aerosol and in embedded states, it could possibly indicate a compaction 
of the aggregates by the pellet pressing, or, since we have already noted that these spectra are 
also influenced by size effects, an aggregate size reduction by the thorough mixing with the 
matrix powder. The size of the aggregates of course could always be influenced by the 
preparation technique. Unfortunately, an inspection of the morphology of the embedded 
aggregates is not possible and a treatment of size effects is beyond the capabilities 
of this theory. For the other samples, which have larger grains and therefore show less 
clumping, we have not observed similar effects. 

\section{Conclusions}
By comparison with experimental spectra, we have demonstrated the usefulness 
of the DFF approach for simulating the infrared extinction spectra of submicron-sized 
particulates composed of oxide and silicate minerals. Predefined DFFs with certain 
characteristics can be used to achieve a satisfactory prediction for the shape 
dependence of the measured infrared band profiles. Good agreement between experimental 
and calculated band profiles was achieved by using 
\begin{enumerate}
\item 
DFFs calculated for Gaussian random spheres with a surface modulation of 
$\sigma$$\sim$0.3 in the case of particulates with irregular grain shapes. 
The main characteristics of these DFFs is a strong and relatively sharp peak at L$\sim$0.1. 
\item
DFFs calculated for aggregates of spheres with a fractal dimension D$_f$=2.4 or 
D$_f$=1.8 in the case of particulates with roundish grain shapes. These DFFs 
peak at L=0.33-0.45 and also have a component at very low L values, which is 
required to fit the experimental band profiles and seems to be correlated 
with the agglomeration state in the particulate. 
\end{enumerate}
Our results suggest that DFF models, such as the ones for GRS with $\sigma$$\sim$0.3 and 
fractal aggregates with D$_f$=2.4 or D$_f$=1.8, would reflect morphological properties 
of real particles significantly better than CDE or DHS models, so would be useful 
in the simulation of cosmic-dust spectra. This could lead to better predictions 
of infrared band profiles or at least to a valid estimate of possible band variations. 
A necessary condition is that grain sizes that are small compared to the wavelength can 
be assumed to dominate the spectrum. 

We recognized that the averaging of spectra obtained by the DFF model for the individual 
crystallographic orientations of an anisotropic material is incorrect and leads in 
some cases to significantly wrong band profiles. This is a problem of all comparable 
theories and can only be overcome by exact simulations, such as a full DDA treatment, 
which however lacks the statistical approach and is difficult to implement in 
comprehensive astronomical spectroscopic simulations. 
Therefore, currently a safer way may be to rely on aerosol-measured spectra.


\begin{acknowledgements}
This work was supported by Deutsche Forschungsgemeinschaft under grant Mu 1164/6-1. 
Special thanks go to W. Teuschel and G. Born for their help with the experiments. We 
are grateful to Prof. C. Koike for providing the Marusu forsterite sample. Furthermore, 
we thank the Elektronenmikroskopisches Zentrum of the medical faculty of the 
Friedrich Schiller University Jena for help with the SEM imaging. H.M. acknowledges 
support by the International Space Science Institute (ISSI) in Bern, Switzerland 
(``Exozodiacal Dust Disks and Darwin'' working group, http://www.issibern.ch/teams/exodust/)"
\end{acknowledgements}


\end{document}